\documentclass{aa}
\usepackage{txfonts}
\usepackage{graphicx}

\begin{document}

\title{The AMIGA sample of isolated galaxies}
\subtitle{III. IRAS data and infrared diagnostics\thanks{Full 
Tables~\ref{tab_fluxes} and  
\ref{tab_lfir} are available in electronic form
at the CDS via anonymous ftp to {\tt cdsarc.u-strasbg.fr (130.79.128.5)} 
or via {\tt http://cdsweb.u-strasbg.fr/cgi-bin/qcat?J/A+A/vvv/ppp} and from
{\tt http://www.iaa.es/AMIGA.html}.}}
%
%
\author{ U. Lisenfeld\inst{1,2}
\and
L. Verdes-Montenegro\inst{2} 
\and 
J. Sulentic\inst{3} 
\and 
S. Leon\inst{4}
\and
D. Espada\inst{2}
\and
G. Bergond\inst{2,5}
\and
E. Garc\'\i a\inst{2}
\and
J. Sabater\inst{2}
\and
J.D. Santander\,-Vela\inst{2}
\and 
S. Verley\inst{2,6,7}
}
\institute{Departamento de F\'\i sica Te\'orica y del Cosmos, 
Facultad de Ciencias, Universidad de Granada, Spain\\ 
\email{ute@ugr.es}
\and
Instituto de Astrof\'{\i}sica de Andaluc\'{\i}a (IAA/CSIC),
Apdo. 3004, 18080 Granada, Spain
\and
Department of Astronomy, University of Alabama, Tuscaloosa, USA
\and
Instituto de Radioastronom\'\i a Milim\'etrica (IRAM), Avda.
Divina Pastora 7, local 20, 18012 Granada, Spain
\and  GEPI/CAI, Observatoire de Paris, 77 avenue Denfert-Rochereau,
  75014 Paris, France
\and LERMA -- Observatoire de Paris, 61 avenue de l'Observatoire, 75014 Paris, France
\and INAF-Osservatorio Astrofisico di Arcetri, Largo E. Fermi 5, 50125 Firenze, Italy
}
\date{Received ; accepted }
\titlerunning{The AMIGA sample of isolated galaxies III}

\authorrunning{Lisenfeld et al.}

  \abstract
   {}
  {We describe the mid- (MIR) and far- (FIR) infrared
properties of a large ($\sim$1000) sample of the most 
isolated galaxies in the local Universe. This sample is 
intended as a ``nurture-free'' zero point against which more
environmentally influenced samples can be compared. } 
   {We reprocess IRAS MIR/FIR survey data using the
ADDSCAN/SCANPI utility for 1030 out of 1050 galaxies from the
Catalogue of Isolated Galaxies (CIG) as part of the AMIGA project.
We focus on diagnostics (FIR luminosity
$L_{\rm FIR}$, $R=\log(L_{\rm FIR}/L_{B})$ and IRAS
colours) thought to be sensitive to effects of environment or interaction.}
   {The distribution of $\log(L_{\rm FIR})$ is sharply peaked from 
9.0--10.5 with very few ($<$2\%) galaxies above 10.5. Review of available
optical images 
of the  most FIR luminous galaxies finds the
majority  to be, likely, interacting systems missed in our earlier
morphological reevaluation. 
The optically normalised luminosity
diagnostic $R= \log(L_{\rm FIR}/L_{B})$ shows a distribution
sharply peaked between 0.0 and $-$1.0. These results were compared to  the
magnitude limited CfA sample that was selected without environmental
discrimination.
This  modestly (e.g. compared to cluster, binary galaxy and compact group
samples) environmentally affected sample shows significantly higher
mean $\log(L_{\rm FIR})$ and $R$, whereas the mean $\log(L_{B})$ is the same.
Our sample shows a strong $L_{\rm FIR}$\ vs. $L_{B}$\ correlation,
with a slope steeper than one ($L_{\rm FIR} \propto L_{B}^{1.41}$). 
Interacting galaxies were found above this
correlation, showing an enhancement in $L_{\rm FIR}$.
With respect to the IRAS colours, we found
higher $F_{\rm 60}/F_{\rm 100}$ value for ellipticals and
late-type galaxies than for spirals, indicating a higher dust temperature.
The mean value of $F_{\rm 60}/F_{\rm 100}$ was found to be lower than
for interacting samples from the literature.}
   {The results indicate that the FIR emission is a variable enhanced by
interaction, and that our sample probably shows the lowest possible
mean value.  This attests to the utility of our sample for defining
a nurture-free zero point.}

\maketitle

\section{Introduction}

Although it is widely accepted that galaxy interactions {\it can}
stimulate secular evolutionary effects in galaxies (e.g. enhanced star
formation, morphological peculiarities including transitions to earlier
type, active nuclei) (e.g. Sulentic \cite{sulentic76}; Hernquist 
\cite{hernquist89}; Xu \& Sulentic \cite{xu91})
there are still many open questions.
Studies aimed at quantifying the level of interaction enhancement have
even produced contradictory results; e.g. some studies of interacting
pairs find a clear star formation enhancement (Bushouse et al. 
\cite{bushouse87}; Bushouse \cite{bushouse88}) while
others find only a marginal increase (Bergvall et al. \cite{bergvall03}). 
Much of this
uncertainty reflects the lack of a statistically useful baseline. What
is the amplitude and dispersion in a given galaxy property that can be
ascribed to ``nature''? A definition of ``isolated galaxy''
is needed before one can properly assess the history and properties of
non-isolated ones.  This has motivated us to build a well-defined and
statistically significant sample of isolated galaxies to serve as a
control sample for the study of galaxies in denser environments.

The AMIGA project (Analysis of the interstellar Medium of Isolated
GAlaxies) involves the
identification and study of a statistically significant sample of the
most isolated galaxies in the local Universe. Our goal is to quantify
the properties of different phases of the interstellar medium in these
galaxies which are likely to be least affected by their external
environment. We adopted the Catalogue of Isolated Galaxies (CIG:
Karachentseva \cite{karachentseva73}; Karachentseva et al. 
\cite{karachentseva86}), including  1051 objects, as a base sample. 
All CIG galaxies are part of the Catalogue of Galaxies and Clusters of
Galaxies providing reasonably uniform apparent magnitude measures
(CGCG: Zwicky et al. \cite{zwicky61})  with $m_{\rm pg} <$ 15.7 and
$\delta > -3 \deg$. Redshifts are now  virtually complete for this
sample with only one of the  compiled objects recognised as a Galactic
source (CIG 781  $\equiv$ Pal 15; Nilson \cite{nilson73}) reducing the
working sample to $n =$ 1050 objects.
AMIGA is compiling data that will characterise 
all phases of the ISM:
blue magnitude, mid- and far-infrared, H$\alpha$, 
and radio continuum fluxes, as well as the
emission of the atomic gas (HI) and
of carbon monoxide (CO), as a tracer of the molecular gas.
The data are being released and periodically updated at
{\tt http://www.iaa.es/AMIGA.html}.

Previous AMIGA papers evaluated, refined and improved the sample in
different ways including:  1) revised positions (Leon \&
Verdes-Montenegro \cite{leon03}), 2) sample redefinition, magnitude correction
and full-sample  analysis of the Optical Luminosity Function (OLF) 
(Verdes-Montenegro et al. \cite{verdes05}: Paper~I)
and 3) morphological revision and type-specific OLF analysis (Sulentic
et al. \cite{sulentic06}: Paper~II). 
In the present paper we  analyse basic mid-
(MIR) and far-infrared (FIR) properties using data from the IRAS
survey (Sects.~\ref{addscan} and \ref{data}). In Sect.~\ref{section4} 
of the paper we present the FIR luminosity function followed by
consideration of various MIR and FIR diagnostics that have been used in
the past to quantify the effects of environment. In Sect.~\ref{section5} 
we discuss the results and compare them to other studies.
Future papers will present a quantification of the 
isolation condition
and the analysis of the radio continuum, H$\alpha$, CO and HI data.

\section{ADDSCAN/SCANPI analysis of the IRAS data}
\label{addscan}
We present co-added ADDSCAN/SCANPI derived fluxes or upper limits for
1030 AMIGA galaxies. The remaining 20 galaxies in our sample were not
covered by the IRAS survey.  Previous studies involving CIG galaxies
worked with smaller subsamples and, in most cases, used IRAS data from
the  Point Source (PSC) and Faint Source Catalogues (FSC).  A subsample
of 476 CIG galaxies with redshifts and PSC fluxes were used as a
control sample for a study of FIR emission from isolated pairs (Xu \&
Sulentic \cite{xu91}, hereafter XS91). Verdes-Montenegro et al. 
(\cite{verdes98}) constructed a reference
sample of 68 CIG galaxies with redshift and blue luminosity
distributions matching their target set of Hickson (\cite{hickson82}) 
compact groups. Hernandez-Toledo et al. (\cite{hernandez01}) 
obtained SCANPI data for 465 CIG
galaxies (those with available redshift data) in order to use them as a
reference in a study of galaxy pairs. FIR data for the CIG galaxies
were however not published in that paper.

IRAS PSC and FSC data exist for only about half of the galaxies in our
sample motivating us to ADDSCAN/SCANPI reprocess our entire sample.  We
used the revised positions from Leon \& Verdes-Montenegro (\cite{leon03}) 
which have a precision of 0\farcs5. ADDSCAN/SCANPI,
a utility provided by the Infrared Processing and Analysis Center (IPAC)
({\tt http://scanpi.ipac.caltech.edu:9000/}), is a one-dimensional tool
that  coadds calibrated IRAS survey data. It makes use of all scans
that passed over a specific position and produces a scan spectrum along
the average scan direction. It is 3--5 times more sensitive than IRAS
PSC since it combines all survey data (Helou et al. \cite{helou88}) and is
therefore more suitable for detection of the total flux from slightly
extended objects.  Our sample was well suited for ADDSCAN/SCANPI processing
because:  i) confusion is minimised since our sample galaxies were
selected with an isolation criterion and ii) the galaxies are small
enough to permit derivation of reliable fluxes.  An analysis of the
IRAS Bright Galaxy Sample (BGS) with ADDSCAN/SCANPI (Sanders et al.
\cite{sanders03}) found that missed flux  became important only for
optical sizes larger than 25\arcmin. About 97\% of the galaxies in our 
sample are smaller than  4\arcmin.

ADDSCAN/SCANPI derives four different flux estimators: 
a) {\it Peak}: maximum flux density within the signal range 
specified, b) {\it fnu}$(z)$: total flux density estimated from
integration of the averaged scan between the zero crossings,
c) {\it Templ}: flux density estimated from the best-fitting 
point source template and d)  {\it fnu}$(t)$:  total flux density  
estimated from integration of the averaged scan between fixed 
points defining an integration range. We adopted the default SCANPI 
ranges (corresponding to the nominal IRAS detector size)
$[-2',+2']$, $[-2',+2']$, $[-2\farcm5,+2\farcm5]$ and $[-4',+4']$ at 
12, 25, 60 and 100~$\mu$m, respectively. We used the median as the 
most robust combination of scans and followed IPAC 
indications in order to choose the best flux density  from
among the estimators for each galaxy. We first flagged as detected 
those galaxies with a $S/N > 3$. We visually confirmed all cases 
and found some errors produced by bright stars in the field or 
baseline corruption from noise or cirrus. 
\begin{table}
\caption{Detection rates and point vs. extended source numbers
for the CIG IRAS sample ($n = 1030$).}
\begin{tabular}{cccccc}
\hline
\hline
$\lambda$ & Threshold& Detections& Detection rate   & Extended & Point \\
\hline
12 &  3$\sigma$ &  180   &   17\%   & 40&       141  \\
   &   5$\sigma$ &   94   &   9\%    & 37&  57 \\
25 &   3$\sigma$ &  245 &   24\%   & 57&       188  \\
   &   5$\sigma$ &   158  &   15\%    & 53&  105 \\
60 &  3$\sigma$ &  729 &   71\%   & 84&       645  \\
   &   5$\sigma$ &   591  &   57\%    & 82&  509 \\
100&   3$\sigma$ &  673 &   65\%   & 37&       636  \\
   &   5$\sigma$ &   526  &   51\%    & 36&  491 \\
\hline
\label{scanpi_stat}
\end{tabular}
\end{table}
\begin{figure}
\resizebox{1.0\hsize}{!}{\includegraphics{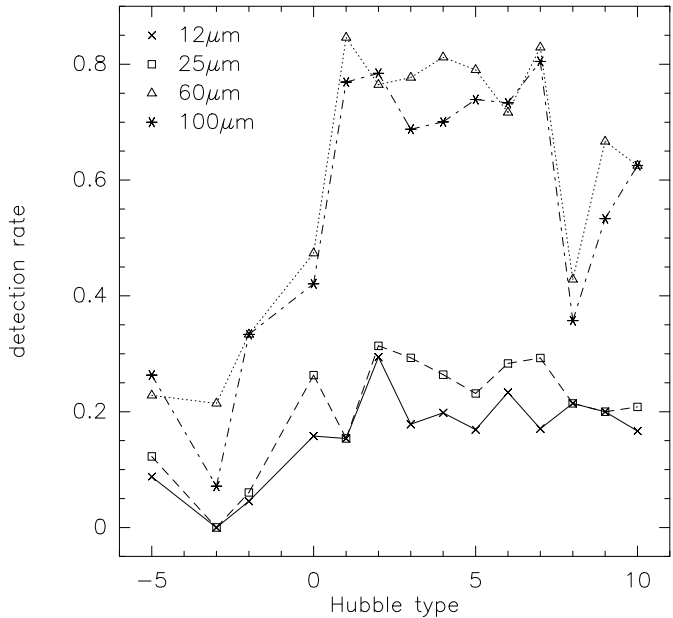}}

\caption{Detection rate at 12, 25, 60 and
 100~$\mu$m  as a function of Hubble type.}
\label{det_rate_hubble}
\end{figure}

Table~\ref{scanpi_stat} summarises the number of detected sources at each 
IRAS band. For completeness, we also include the corresponding numbers
for a detection threshold of 5 times the rms noise level, which
is the limit used in the data analysis carried out in this
paper (see Sect.~\ref{sample_definition}). 
Figure~\ref{det_rate_hubble} shows the detection rate 
(at a 3$\sigma$ detection threshold) at the four IRAS 
wavelengths as a function of Hubble type.
The MIR-FIR detection rates show a minimum for early-type
galaxies gradually increasing from 10--20\% to 20--80\% for late-type spirals. 
We see a decline to 20--60\% beyond type Sd ($T=7$) probably reflecting an 
increasing dwarf galaxy population with low  dust masses.

\begin{figure*}
\includegraphics{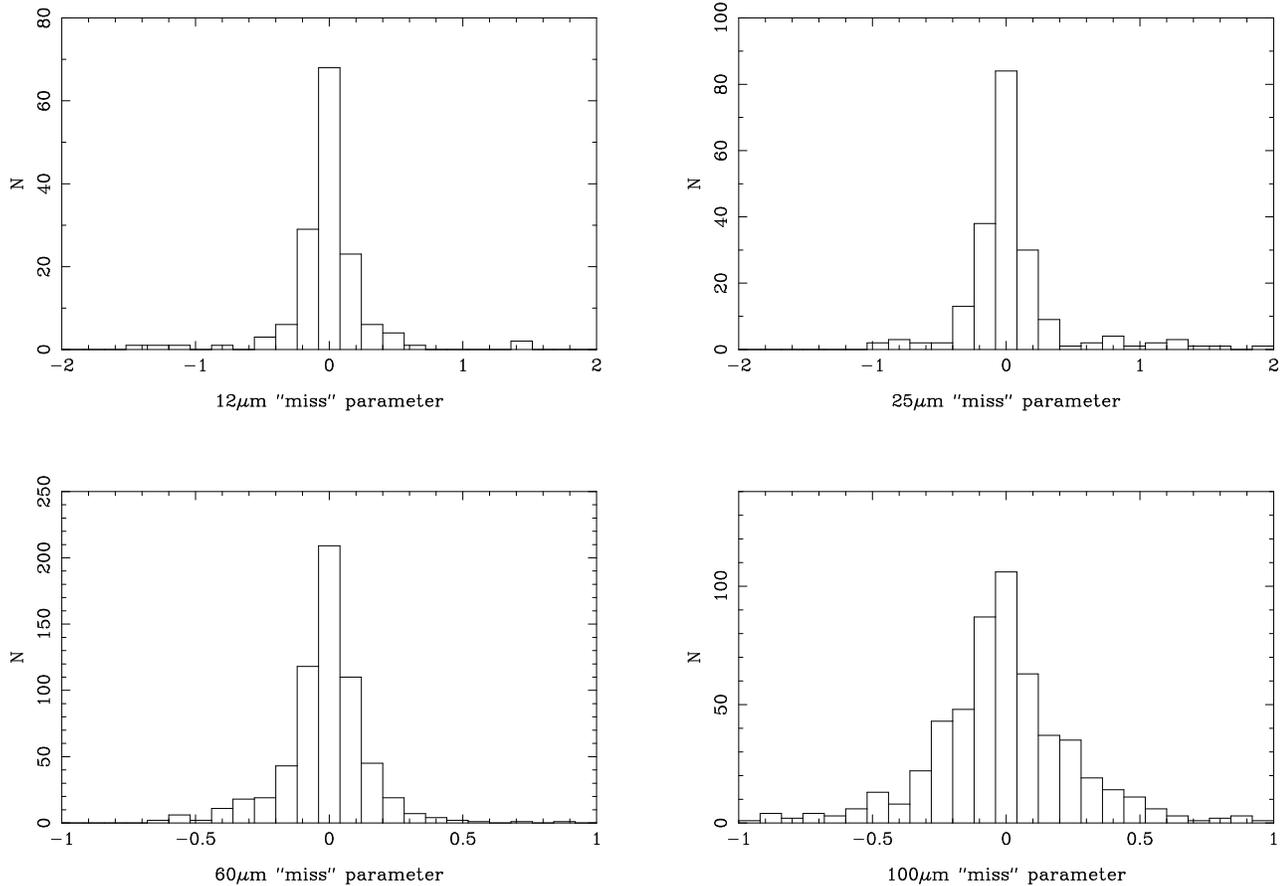}
\caption{Histogram of the ``miss'' parameter (offset in arcmin between the 
galaxy position and the position of the
signal peak) 
for each IRAS band.}
\label{fig-miss.ps}
\end{figure*}

Figure~\ref{fig-miss.ps}  plots the ``miss'' parameter which
is the offset in arcmin between the galaxy position and the position of the
signal peak along the average scan direction. This parameter is used as
the primary measure of source identification. The majority of sources
cluster around zero offset with  the largest deviations occurring 
at 12 and  25~$\mu$m  because: 1) the resolution is higher and 2) the $S/N$ 
is usually lower than at longer wavelengths. The standard deviations of the 
``miss'' parameter are 18\arcsec, 24\arcsec, 14\arcsec\ and 28\arcsec, 
respectively, for 12, 25, 60 and 100~$\mu$m. This is a factor of 
$\frac{1}{2}$ to $\frac{1}{6}$ of the 
nominal FWHM of the IRAS detectors (0\farcm77, 0\farcm78, 1\farcm44, and 
2\farcm94 at 12, 25, 60 and 100~$\mu$m respectively).
This scatter is reasonable when one allows for the fact that most of these 
galaxies are not very infrared (IR) bright so that determination of the source
centroid depends somewhat on the $S/N$ of the measurement.

In the next step we used two different tests to decide whether a
detected source was  extended or pointlike with respect to the IRAS
beam.  In Test 1 we considered as extended those galaxies where the
signal width was greater than the expected width for a point source.
We used both {\it w25} and  {\it w50} (width of the signal in
arcminutes at 25\% and 50\% peak) for this comparison. We compared our
measures to the widths of point sources (Sanders et al. \cite{sanders03}) where
{\it w25psf} and {\it w50psf} were 1\farcm40, 1\farcm38, 2\farcm06, 4\farcm32 
and 1\farcm04, 1\farcm00, 1\farcm52, 3\farcm22 at 12, 25, 60, 100~$\mu$m, 
respectively. In
Test 2 we classified as extended those sources where the integrated
flux, {\it fnu}$(z)$ was substantially larger than the peak flux adopting
the condition {\it fnu}$(z) - {\it Peak} >  5\sigma$ as a threshold
criterion for an extended source. The percentages of galaxies showing
conflicting classifications in the two tests were 9, 17, 23 and 18\% at
12, 25, 60 and 100~$\mu$m, respectively. We revised these cases
interactively and found the differences were most often due to baseline
corruption by noise and/or cirrus. Table~\ref{scanpi_stat} lists the
number of sources classified as point  or extended for each IRAS band.
The 5$\sigma$ cutoff reduces, compared to the 3$\sigma$ cutoff,
 mainly the number of detected point sources
and leaves the number of detected extended sources almost unchanged.
The reason is that sources classified as extended  have in most cases $S/N > 5$.
Once the size of a source was decided  we could choose a flux
estimator following guidelines given by IPAC.  For point sources three
cases were considered:  a) if {\it Templ} $>$ 20~Jy we used {\it Peak},
b) if {\it Templ} $<$ 1~Jy ($<$ 2~Jy at 100~$\mu$m) we used  {\it Templ}
and c) if  1~Jy $<$ {\it Templ} $<$ 20~Jy (2~Jy $<$ {\it Templ} $<$ 20
Jy at 100~$\mu$m) we used  {\it Templ} if  {\it Peak} and  {\it fnu}$(t)$
agreed within 3$\sigma$. Otherwise  we visually determined the best
estimator of the total flux density. In the case of extended sources we
used {\it fnu}$(z)$ when $S/N$ $>$ 10 and {\it fnu}$(t)$ for the rest.

\section{The data\label{section3}}
\label{data}

\begin{table*}
      \caption{FIR flux densities$^{1}$.}
\begin{tabular}{rrrrrrrrrrrrrrrrr}
\hline
\hline
   & 
\multicolumn{4}{c}{12~$\mu$m}&
\multicolumn{4}{c}{25~$\mu$m}&
\multicolumn{4}{c}{60~$\mu$m}&
\multicolumn{4}{c}{100~$\mu$m}\\
 (1) &
\multicolumn{1}{c}{(2)}& 
 \multicolumn{1}{c}{(3)}&
 \multicolumn{1}{c}{(4)}& 
 \multicolumn{1}{c}{(5)}&
 \multicolumn{1}{c}{(6)}&    
 \multicolumn{1}{c}{(7)}&  
 \multicolumn{1}{c}{(8)}& 
 \multicolumn{1}{c}{(9)}&
 \multicolumn{1}{c}{(10)}&    
 \multicolumn{1}{c}{(11)}&  
 \multicolumn{1}{c}{(12)}& 
 \multicolumn{1}{c}{(13)}&
 \multicolumn{1}{c}{(14)}&    
 \multicolumn{1}{c}{(15)}&  
 \multicolumn{1}{c}{(16)}& 
 \multicolumn{1}{c}{(17)}\\
 CIG &
\multicolumn{1}{c}{$F_{12}$}& 
 \multicolumn{1}{c}{rms}&
 \multicolumn{1}{c}{M}& 
 \multicolumn{1}{c}{E}&
 \multicolumn{1}{c}{$F_{25}$}&    
 \multicolumn{1}{c}{rms}&  
 \multicolumn{1}{c}{M}& 
 \multicolumn{1}{c}{E}&
 \multicolumn{1}{c}{$F_{60}$}&    
 \multicolumn{1}{c}{rms}&  
 \multicolumn{1}{c}{M}& 
 \multicolumn{1}{c}{E}&
 \multicolumn{1}{c}{$F_{100}$}&    
 \multicolumn{1}{c}{rms}&  
 \multicolumn{1}{c}{M}& 
 \multicolumn{1}{c}{E}\\
 &
\multicolumn{1}{c}{(Jy)}& 
 \multicolumn{1}{c}{(Jy)}&
 \multicolumn{1}{c}{}& 
 \multicolumn{1}{c}{}&
 \multicolumn{1}{c}{(Jy)}&    
 \multicolumn{1}{c}{(Jy)}&  
 \multicolumn{1}{c}{}& 
 \multicolumn{1}{c}{}&
 \multicolumn{1}{c}{(Jy)}&    
 \multicolumn{1}{c}{(Jy)}&  
 \multicolumn{1}{c}{}& 
 \multicolumn{1}{c}{}&
 \multicolumn{1}{c}{(Jy)}&    
 \multicolumn{1}{c}{(Jy)}&  
 \multicolumn{1}{c}{}& 
 \multicolumn{1}{c}{}\\
\hline
    1& $<$  0.07& 0.02&    5&  & $<$  0.23& 0.08&    5&  &      0.86& 0.07&    1&n &    2.87& 0.16&    2&n\\
    2& $<$  0.09& 0.03&    5&  & $<$  0.11& 0.04&    5&  &      0.21& 0.05&    1&n &    0.74& 0.21&    1&n\\
    3&      0.06& 0.02&    1&n & $<$  0.08& 0.03&    5&  &      0.19& 0.03&    1&n &    0.43& 0.07&    1&n\\
    4&      0.66& 0.03&    4&y &      0.61& 0.03&    4&y &      5.19& 0.05&    4&y &   16.78& 0.12&    4&y\\
    5& $<$  0.13& 0.04&    5&  &      0.12& 0.04&    2&n &      0.25& 0.04&    1&n &    0.76& 0.14&    1&n\\
\ldots & \ldots &\ldots & \ldots &\ldots & \ldots &\ldots & \ldots &\ldots & \ldots &\ldots & \ldots &\ldots & \ldots &\ldots & \ldots &\ldots\\
\hline
\end{tabular}


The Table format is: {\it Column 1}: CIG number.
{\it Column 2}: Flux density at 12~$\mu$m, calculated as explained in 
Sect.~\ref{addscan}. Upper limits are preceded by a ``$<$'' sign. 
A  3$\sigma$ value has been adopted for the upper limits,  except for 
CIG 397 where the 12~$\mu$m scan presents
confusion with a close star and 20\% of the peak of the emission has 
been adopted as an upper limit.
{\it Column 3}: rms noise of the data at 12~$\mu$m.
{\it Column 4}: Method used to derive the flux densities given in column (2). 
Codes 1 to 4 correspond to the following flux estimators:
1 =  {\it Templ}, 2 = {\it Peak}, 3 =  {\it fnu}$(t)$, 4 = {\it fnu}$(z)$. 
Code 5 corresponds to upper limits obtained as 3$\sigma$. 
Code 6 is reserved for some particular cases: CIG 397 (see above) 
and nine galaxies included in the catalogue of large
optical galaxies of Rice et al. (\cite{rice88}) (CIG~105, 197, 324, 347, 461, 
469, 523, 559 and 610) where we have used the values given in their catalogue 
(see also Sect. \ref{comparison}). 
{\it Column 5}: Detected galaxies are flagged with ``y''
if they have been classified as extended, and with ``n'' when 
classified as point sources.
{\it Column 6--9}: The same as column 2--5 for   25~$\mu$m.
{\it Column 10--13}: The same as column 2--5 for   60~$\mu$m.
{\it Column 14--17}: The same as column 2--5 for  100~$\mu$m.\\
$^{1}$ The full table is available in electronic form
at the CDS  
and from {\tt http://www.iaa.es/AMIGA.html}.
\label{tab_fluxes}
\end{table*}
Table \ref{tab_fluxes} lists $\lambda$ 12, 25, 60 and 100~$\mu$m derived fluxes
obtained using procedures explained in Sect.~\ref{addscan}. We also tabulate
some related parameters, as detailed in the notes to the table.

\subsection{Comparison to other IRAS catalogues}
\label{comparison}

We compared IRAS fluxes obtained with SCANPI to data available from the
IPAC archives and in the literature. We retrieved data from the Faint Source 
Catalogue (FSC) and the Point Source Catalogue (PSC) from the
IRAS database through the GATOR service
({\tt http://irsa.ipac.caltech.edu/}).  We found 509 CIG galaxies in the FSC
and  additional data for 15 galaxies in the PSC.
The average
error-weighted ratios $F$(SCANPI)/$F$(FSC+PSC) for galaxies detected
both by SCANPI and in the FSC+PSC are $1.24 \pm 0.50$ $(n=114)$,  $1.16
\pm 0.33$ $(n=153)$,  $1.09 \pm 0.15$ $(n=501)$, $1.05 \pm 0.13$ $(n=407)$ for
12, 25, 60 and 100~$\mu$m.  The average ratio is slightly larger than
one and decreases with increasing wavelength. This indicates that the
flux derivation with SCANPI is able to pick up more flux for extended
objects than FSC/PSC especially at short wavelengths were the IRAS beam
is smaller. There is a large number of galaxies with FSC/PSC tabulated
upper-limits ($n=55$, 70, 9, and 81 for 12, 25, 60 and 100~$\mu$m) that
were replaced by SCANPI detections indicating that the detection rate
has been improved by the reprocessing. Other galaxies
were listed as FSC/PSC detections while SCANPI derived only upper
limits ($n=29$, 21, 5, and 3 for 12, 25, 60 and 100~$\mu$m).  We checked
those cases individually and found that all were weak sources
where either: 1) the automated SCANPI procedure did not confirm a detection 
or 2) we decided, after visual inspection, that $S/N<3$.

\begin{figure*}
\resizebox{0.5\hsize}{!}{\includegraphics[angle=270]{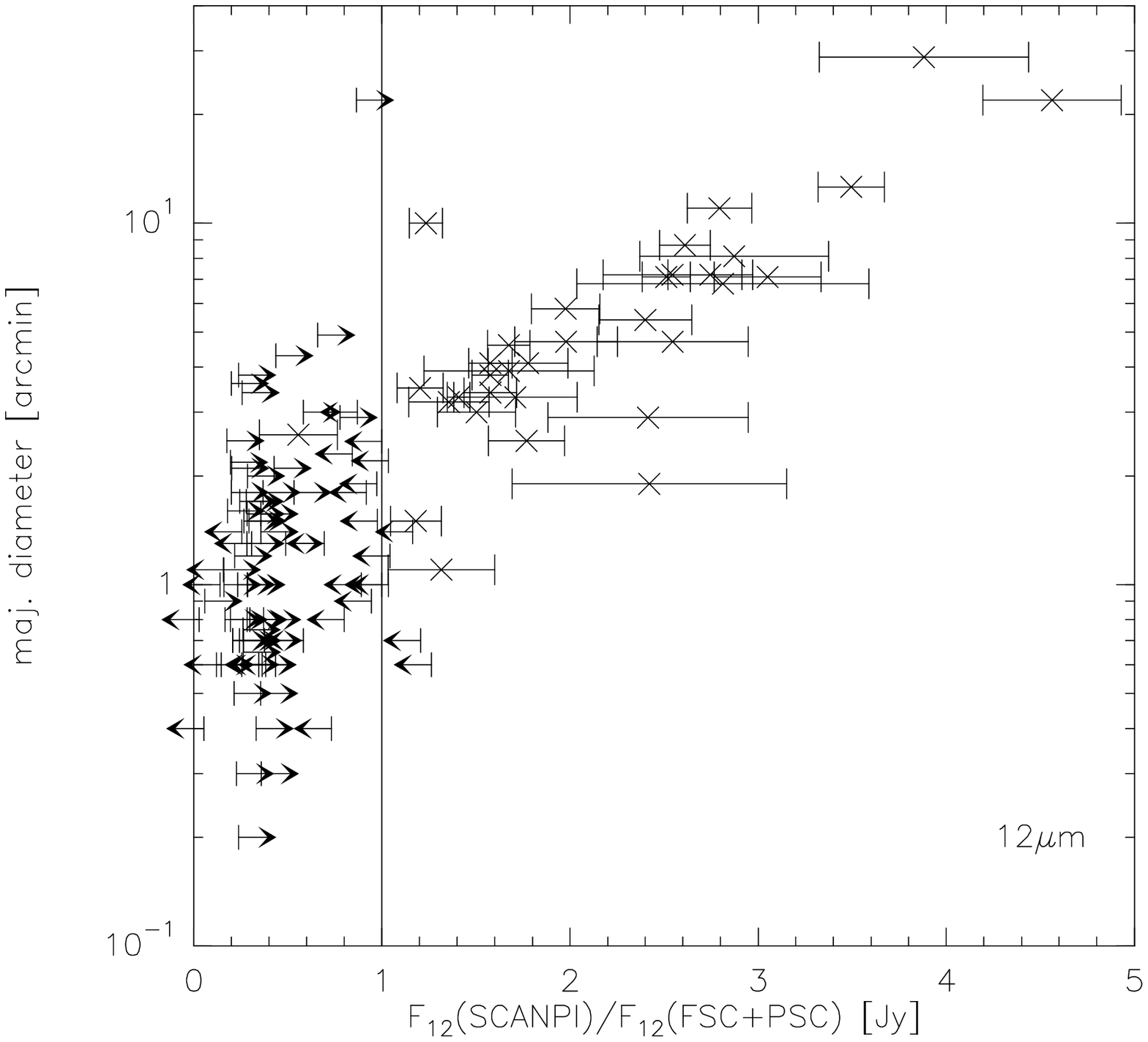}}
\resizebox{0.5\hsize}{!}{\includegraphics[angle=270]{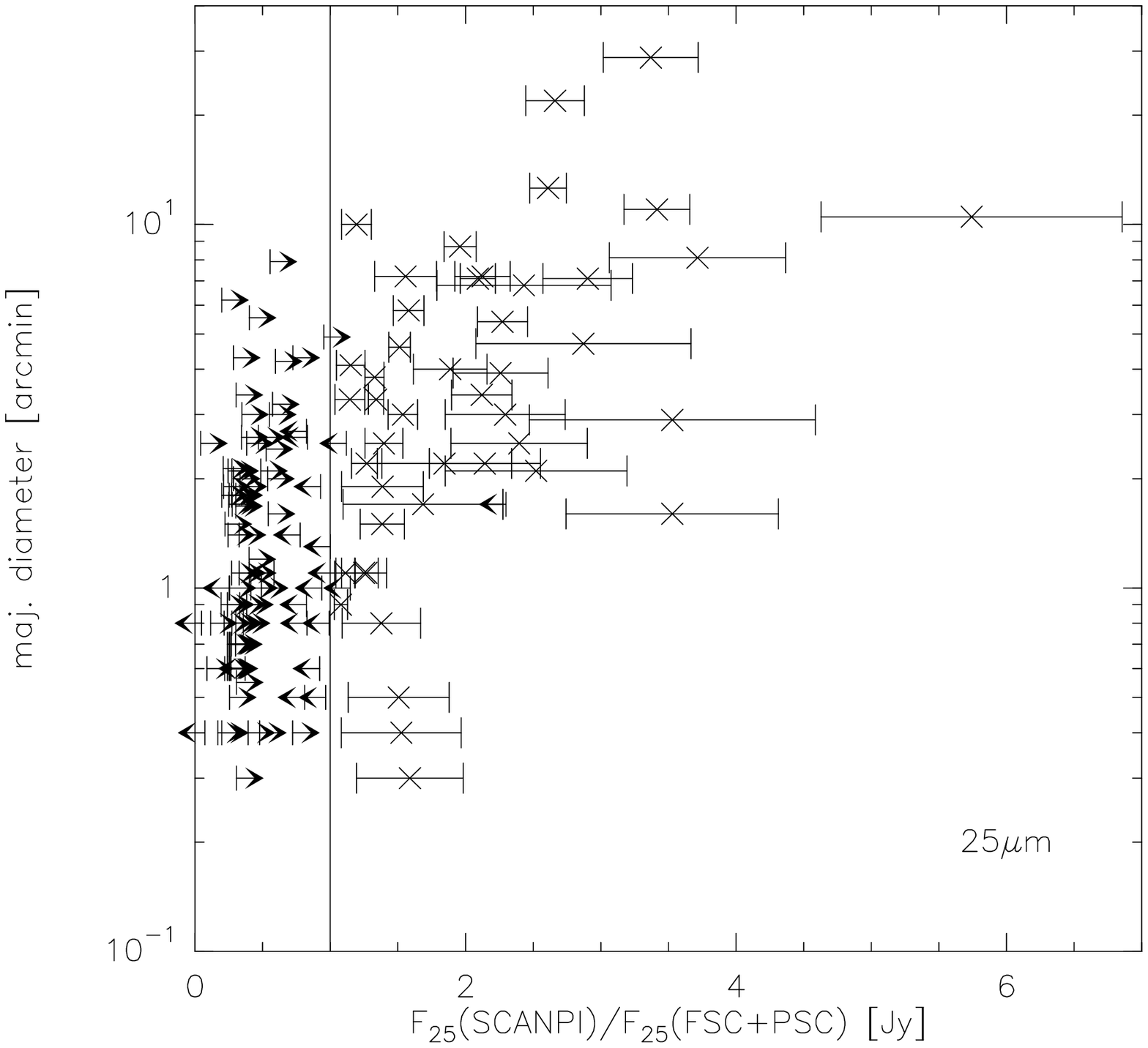}}
\resizebox{0.5\hsize}{!}{\includegraphics[angle=270]{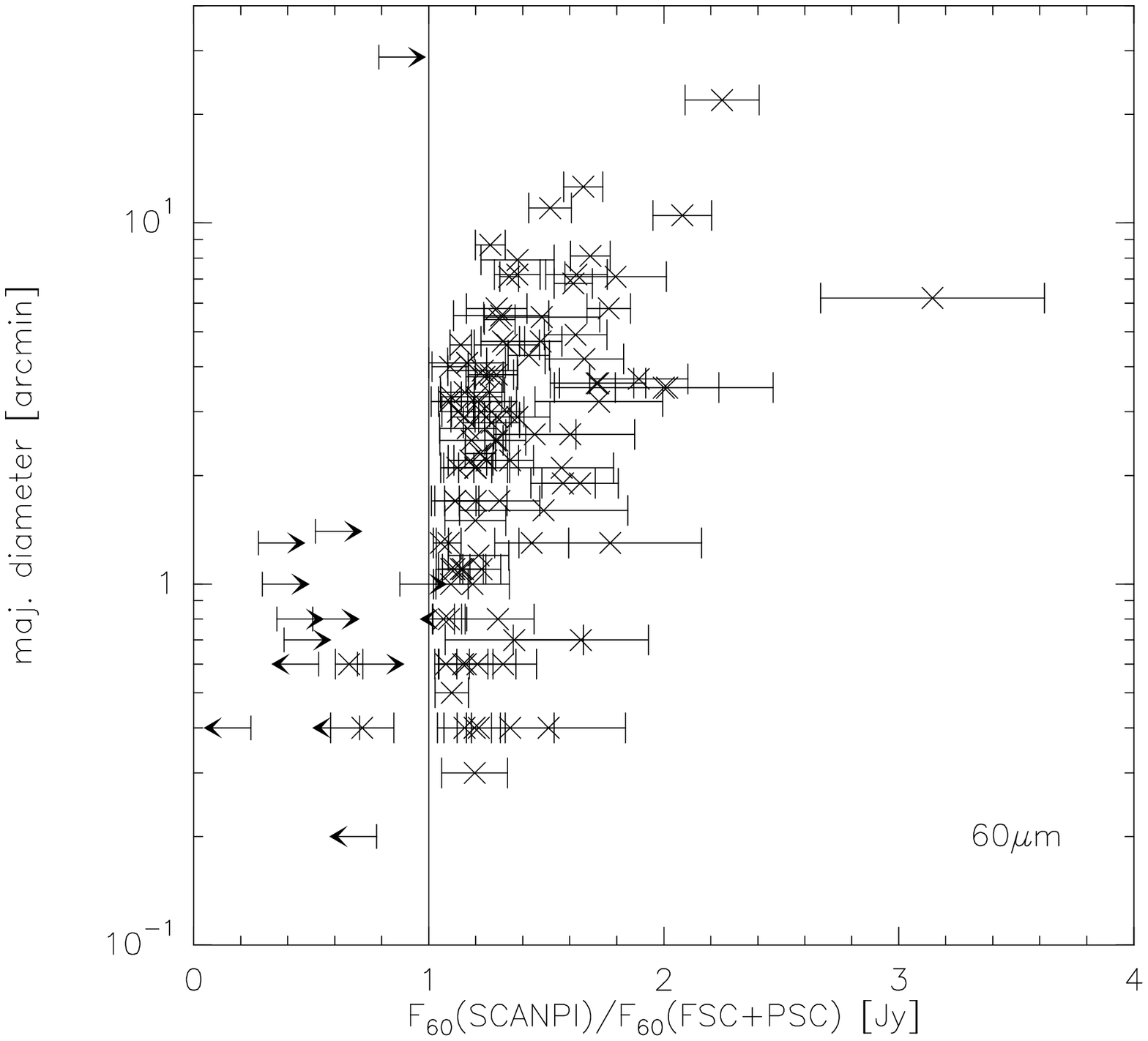}}
\resizebox{0.5\hsize}{!}{\includegraphics[angle=270]{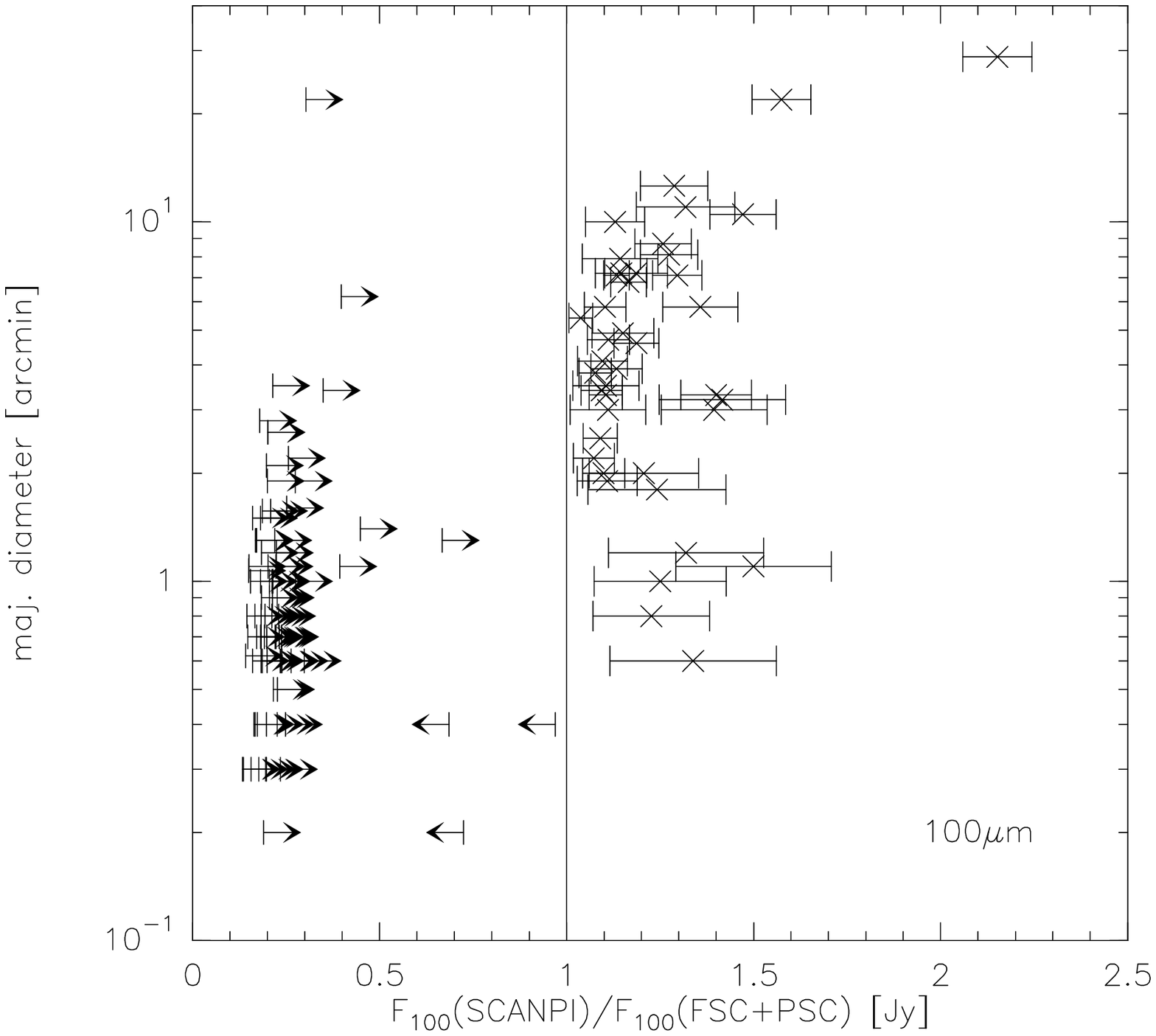}}
\caption{SCANPI/FSC+PSC flux ratio as a function of optical diameter for 
the 4 IRAS wavelengths. Detected galaxies showing consistent fluxes
between SCANPI and FSC+PSC are not included in the figure for clarity.
}
\label{compare_ipac_gator}
\end{figure*}

Figure~\ref{compare_ipac_gator} shows the SCANPI-to-FSC/PSC flux ratio as a
function of optical diameter for each IRAS band. We exclude detections where 
SCANPI and FSC/PSC fluxes agree within the uncertainties (which is the case 
for 82, 107, 397 and 368 galaxies at 12, 25, 60 and 100~$\mu$m, respectively). 
The 12 and 25~$\mu$m plots, and to a lesser extent 60~$\mu$m, show that
$F$(SCANPI)/$F$(FSC+PSC) increases  with optical diameter above about
$1'$--$3'$. This supports our inference that FSC/PSC flux
densities are often underestimated for large galaxies because part of
the flux falls outside the IRAS beam  
and that
SCANPI is able to provide a more realistic flux estimate for these
sources.  There are only three galaxies (CIG~546, 616 and 721) with
$F$(SCANPI)/$F$(FSC+PSC) significantly below one. In the case of CIG 546
($F_{60}$(SCANPI)/$F_{60}$(FSC+PSC) = 0.66) and CIG 721
($F_{60}$(SCANPI)/$F_{60}$(FSC+PSC) = 0.72) the origin of the
difference is unclear since we have no reason to doubt the reliability
of our SCANPI estimates. CIG~616 ($F_{12}$(SCANPI)/$F_{12}$(FSC+PSC) =
0.56) was detected by SCANPI just above a  3$\sigma$ level so that the
flux density has a larger uncertainty. 

We compared 18 galaxies in common with the Bright IRAS Galaxy
sample ($F_{60} > 5.24$~Jy) where flux densities were also derived using 
SCANPI (Sanders et al. \cite{sanders03}). Agreement is better than 6\% for
all sources at 12, 60 and 100~$\mu$m. At 25~$\mu$m there are three sources
(CIG~442, 549 and 1004) where our adopted values exceed those derived
by Sanders et al. (\cite{sanders03}) at the  10--20\% level.  
We think that the discrepancy
arises because some of the flux in these sources extends beyond the
integration range used in deriving {\it fnu}$(t)$ and will therefore be
better estimated using our {\it fnu}$(z)$ values.  

Following IPAC
recommendations we compared the results derived with SCANPI to those
derived from 2D Full Resolution Coadded (FRESCO) images for sources
with optical diameter larger than 2\farcm5 (107 objects). Since
FRESCO images do not have large-scale background removed (they are not
point-source filtered), they provide additional information about
galaxy environments including possible confusion due to nearby stars or
Galactic cirrus. We extracted individual source fluxes from FRESCO images 
using Sextractor (Bertin \& Arnouts \cite{bertin96}).  We extracted fluxes
for CIG galaxies using both 3 and 5$\sigma$ thresholds above the
local background level in order to estimate the effects of background
and particularly Galactic cirrus.  Calibrated FRESCO fluxes for the 4
IRAS bands were compared to the SCANPI fluxes  and we found ratios
$F$(SCANPI)/$F$(FRESCO) close to unity ($1.04 \pm 0.42, 0.98 \pm 0.40,
0.89 \pm 0.33$ and $0.97 \pm 0.44$ at 12, 25, 60 and 100~$\mu$m)
respectively. Scatter was high  but we did not find a trend with
optical diameter that might point towards flux being missed using either
procedure.  More likely contamination from the local foreground
(Galactic emission) is dominating the  flux determination.

Finally, we searched for CIG galaxies included in the catalogue of large 
optical galaxies (Rice et al. \cite{rice88}) and found nine objects (CIG~105,
197, 324, 347, 461, 469, 523, 559 and 610).  In most cases we find reasonable
(within 2$\sigma$) agreement between flux estimates.  There are some
significant discrepancies for the largest objects. The most severe
discrepancy involves the galaxy with largest apparent optical diameter
CIG~610 ($\equiv$ M101, $28\farcm8 \times 26\farcm9$) where SCANPI fluxes
are only 10--35\% of the fluxes given in Rice et al. (\cite{rice88}).
Disagreements of up to a  factor of 2.5 are present for  CIG~197 ($\equiv$ 
NGC~2403, $21\farcm9 \times 12\farcm3$) and CIG~523 ($\equiv$ NGC~4236,
$21.9^\prime \times 7.2^\prime$).  The single discrepancy involving a
smaller galaxy, CIG~105 ($\equiv$ NGC~925,  $10\farcm5\times 5\farcm9$)
finds a SCANPI flux at 12~$\mu$m that is almost a factor two lower than
the flux in Rice et al. (\cite{rice88}). We adopted the flux densities of 
Rice et al. (\cite{rice88}) for all nine galaxies.

\section{Data analysis\label{section4}}
\subsection{Sample definition}
\label{sample_definition}
In the following sections we analyse the FIR emission properties of 
the CIG galaxy sample. In order to do this in a 
statistically meaningful way we focus on the optically 
complete sample described in Paper~I. This 
sample involves galaxies with corrected Zwicky magnitudes in the range 
11.0--15.0 for which we found $<V/V_m> = 0.40$, indicating 80\% completeness.

We include in the present work some changes/upgrades with respect to the 
previously (in Paper~I) defined sample: 
\begin{enumerate}

\item  We include 20 galaxies for which redshift information has become 
available since the publication of Paper~I (the updated redshift list 
is available from {\tt http://www.iaa.es/AMIGA.html}).

\item Morphological revision of the sample, described in Paper~II
identified 32 galaxies that are probably not isolated in the sense that they might involve
isolated interacting pairs and/or multiplets. These galaxies are excluded 
from the most isolated sample and represent part of the AMIGA refinement.
However they provide us with a useful internal comparison sample to test 
the effects of interaction contamination. 

\item We recomputed corrections to the Zwicky magnitudes following
Paper~I but using the revised morphologies from Paper~II. This change 
in individual magnitudes will therefore slightly change the sample involving 
galaxies in the range 11.0--15.0 mag. The present sample shows a value
of  $<V/V_m> = 0.43$ indicating a slightly improved level of completeness
compared to Paper~I.

\item We exclude  two nearby dwarf ellipticals  (CIG~663 $\equiv$ Ursa Minor 
and CIG 802 $\equiv$ Draco) for which we have only IRAS upper limits and very 
low inferred luminosity limits ($\log(L_{\rm FIR}/L_{\sun}) < 3.25$).
\end{enumerate}
 
We are left with a sample of  719  galaxies with IRAS data, and redshift
data  is available for 701 galaxies of them. 
Hereafter we will refer to this sample as the AMIGA (FIR) sample.
We decided to increase the detection threshold to 5 $\sigma$ in order
to make sure that we only consider reliable detections. Thus, with
respect to Table~\ref{tab_fluxes}, we now consider only those fluxes 
as detections where the $S/N$ ratio is above 5, and we use an upper
limit of 5 times the rms noise for values below.
(We chose to leave the 3$\sigma$ detection limit in   Table~\ref{tab_fluxes}
in order to provide the complete data set.)
With this higher threshold, 511 galaxies have a detection at least at one 
wavelength. 
This sample can be cut in
many different ways. Right now we make no restriction in recession
velocity.  This allows us to sample the widest possible luminosity
range. Sources with $V \le 1500$~km\,s$^{-1}$ provide an insight into the IR
emission from local dwarf galaxies that are not included in the
rest/bulk of the sample.  The drawback about including these galaxies
in the sample involves the difficulty in reliably assessing their
isolation properties. 

\subsection{FIR luminosity}
\label{luminosity}

FIR luminosity, $L_{\rm FIR}$, is computed from IRAS measurements
as  $\log(L_{\rm FIR}/L_{\sun}) = \log(FIR) + 2 \log(D) + 19.495$,
where D is distance in Mpc and 
$FIR = 1.26 \times 10^{-14} (2.58F_{60} + F_{100})$ W\,m$^{-2}$
(Helou et al. \cite{helou88}) the flux in the FIR range, with the IRAS 
fluxes at 60 and 100~$\mu$m, 
$F_{60}$ and $F_{100}$. 
$L_{\rm FIR}$ and the distances adopted are listed in Table~\ref{tab_lfir}.

\begin{table}
      \caption{FIR and blue luminosities$^1$.}
\begin{tabular}{rrrrrrrrr}
\hline
\hline
(1) & (2) & (3) & (4)\\
CIG & 
 \multicolumn{1}{c}{Distance}& 
 \multicolumn{1}{c}{$\log(L_{\rm FIR})$} &
 \multicolumn{1}{c}{$\log(L_{B})$}\\
& 
 \multicolumn{1}{c}{(Mpc)}& 
 \multicolumn{1}{c}{($L_{\sun}$)}&
 \multicolumn{1}{c}{($L_{\sun}$)}\\
\hline
    1&  92.2 &10.23 & 10.44\\
    2&  88.7 & $<$ 9.72  & 9.76 \\
    3&  \_     & \_   &  \_  \\
    4&  26.1 &9.91  & 10.17\\
    5& 100.2 &9.75  & 10.07\\
\ldots  & \ldots   & \ldots  &\\
\hline
\end{tabular}

The entries are: 
{\it Column (1)}: CIG number. {\it Column (2)}: Distance in Mpc  from Paper~I.
{\it Column (3)}: FIR luminosity, calculated as described in  
Sect.~\ref{luminosity}. Upper limits are indicated with $<$ in front of the 
value. Galaxies with distances, but without FIR data points (in total: 20 objects) lie in the area 
not covered by IRAS. {\it Column (4)}: Blue luminosity, calculated as 
described in  Sect.~\ref{lfir_and_lb}.\\
$^1$ The full table is available in electronic form
at the CDS 
and from {\tt http://www.iaa.es/AMIGA.html}.
\label{tab_lfir}
   \end{table}

\begin{figure}
\resizebox{1.0\hsize}{!}{\includegraphics{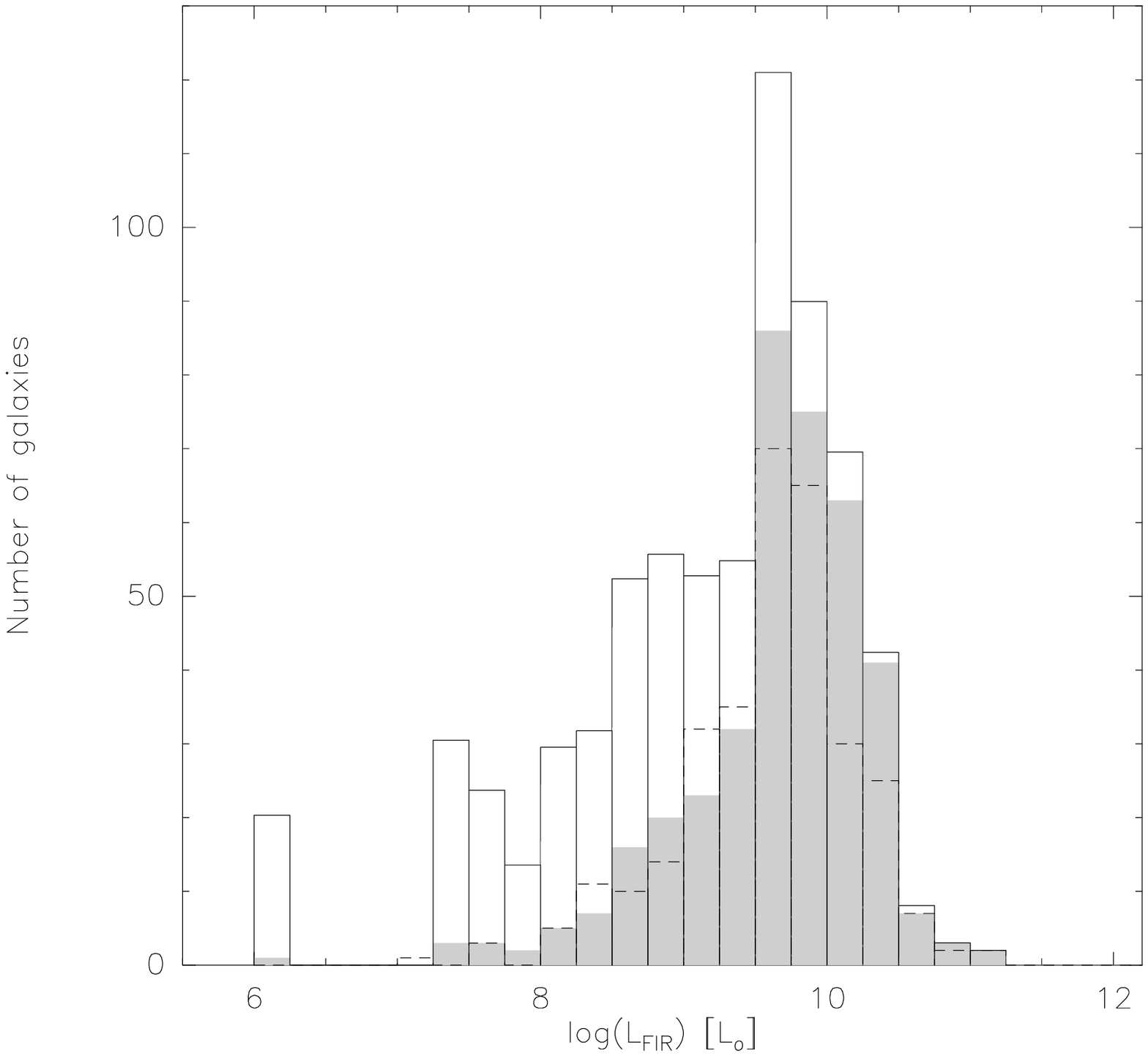}}
\caption{The FIR luminosity distribution of the optically selected sample 
described in Sect.~\ref{sample_definition}. The full line shows the 
distribution calculated with ASURV ($n=701$), the shaded area shows all  
galaxies detected  at both 60 and 100~$\mu$m ($n = 478$), and the  dashed 
line gives the non-detections.}
\label{lfir_histo}
\end{figure}

\begin{table*}
      \caption{Mean and median values of $L_{\rm FIR}$, $L_{B}$, and 
$R=\log(L_{\rm FIR}/L_{B})$.}
\begin{tabular}{lccccccc}
\hline
\hline
(1) & (2) & (3) & (4) & (5) & (6) & (7) & (8)    \\
Sample & $n$ & $<\log(L_{B})>$ & $n_{\rm up}$ & $<\log(L_{\rm FIR})>$ & 
$n$  & $n_{\rm up}$ & $<R>$ \\
 & &   med$(\log(L_{B}))$& & med$(\log(L_{\rm FIR}))$ & & &   med($R$) \\
\hline
Total & 701 & 9.97$\pm$0.02&  312 & 9.15$\pm$0.06 & 719&327 &$-$0.56$\pm$0.03\\
  & & 10.06 &   & 9.74 & & & $-$0.29     \\
S/Im ($T=1$--10) & 616 & 9.98$\pm$0.02 & 248 & 9.26$\pm$0.05 & 634 & 263 & $-$0.49$\pm$0.02 \\
  & & 10.07 &   & 9.76 & & & $-$0.30     \\
\hline
E ($T=-5$)     & 27 & 9.95$\pm$0.06& 21 & 8.83$\pm$0.16&  27 &21  &$-$1.01$\pm$0.12 \\
  & & 10.01 &   & 9.77 & & & $-$0.14     \\
S0 ($T=-2$)    & 36 & 9.82$\pm$0.08& 27 & 8.58$\pm$0.15 & 36 &27  &$-$0.95$\pm$0.09\\
  & & 9.95 &   & 9.65 & & & $-$0.23     \\
S0a ($T=0$)    & 14 & 9.88$\pm$0.07 & 8 & 9.33$\pm$0.16 & 14 & 8  &$-$0.57$\pm$0.15 \\
  & & 9.88 &   & 9.73 & & & $-$0.03     \\
Sa ($T=1$)     & 10 & 9.92$\pm$0.18& 3 & 9.27$\pm$0.20 & 10 & 3  &$-$0.60$\pm$0.10\\
  & & 10.00 &   & 9.64 & & & $-$0.49    \\
Sab ($T=2$)    & 39 &10.05$\pm$0.05& 15 & 9.39$\pm$0.10 &  39 & 15 &$-$0.62$\pm$0.08\\
  & & 10.00 &  & 9.61 & & & $-$0.29    \\
Sb ($T=3$)     &115 &10.06$\pm$0.04& 42& 9.54$\pm$0.08 &118 & 45 &$-$0.38$\pm$0.04\\
  & & 10.10 &  & 9.90 & & & $-$0.14    \\
Sbc ($T=4$)    &155 &10.10$\pm$0.03& 70 & 9.37$\pm$0.09 &160 & 73 &$-$0.50$\pm$0.03\\
  & & 10.15 &  & 9.82 & &  & $-$0.32    \\
Sc ($T=5$)     &182 &10.12$\pm$0.03& 69 & 9.62$\pm$0.05 &188 & 74 &$-$0.38$\pm$0.02\\
  & & 10.20 &  & 9.91 & & & $-$0.28   \\
Scd ($T=6$)    & 47 & 9.65$\pm$0.08&  19 & 8.89$\pm$0.12 & 47 & 19 &$-$0.54$\pm$0.06\\
  & & 9.77 &   & 9.28 & & & $-$0.35  \\
Sd ($T=7$)     & 34 & 9.58$\pm$0.09&  15& 8.73$\pm$0.15 &  38 & 19 &$-$0.64$\pm$0.06\\
  & & 9.51 &   & 8.95 & & & $-$0.52  \\
Sdm($T=8$)     & 10 & 9.38$\pm$0.21&  7 & 8.40$\pm$0.19 & 10 & 7  &$-$0.55$\pm$0.05\\
  & & 9.05 &   & 8.49 & & & $-$0.50 \\
Sm ($T=9$)     &  9 & 9.11$\pm$0.34&  5 & 7.98$\pm$0.27 & 9  & 5  &$-$0.58$\pm$0.05\\
  & & 9.07 &   & 8.56 & & & $-$0.61\\
Im($T=10$)     & 15 & 9.01$\pm$0.21&  3 & 8.30$\pm$0.34 & 15 & 3  &$-$0.58$\pm$0.13\\
  & & 9.04 &   & 8.70 & & & $-$0.47\\
\hline
Interacting& 14 & 9.99$\pm$0.11& 2   & 9.87$\pm$0.20 &  14 & 2  &$-$0.06$\pm$0.08 \\
  & & 9.98 &   &  10.02 & & & $-$0.11\\
\hline
\end{tabular}

The entries are: 
{\it Column 1}:  Subsample considered. All subsamples are selected from the 
optically complete, magnitude limited subsample. The interacting subsample 
consists of the galaxies from the CIG excluded in Paper~II 
(see Sect. \ref{sample_definition}).
{\it  Column 2}:  Total number of galaxies with velocity and IRAS data in the 
subsample.
{\it Column 3}:  First row: Mean value of $L_{B}$.  Second row: Median value 
of $L_{B}$.
{\it Column 4}:  Number of upper limits in FIR (at 60 or 100~$\mu$m). 
{\it Column 5}:  First row: Mean value of $L_{\rm FIR}$, using the 
Kaplan-Maier estimator from ASURV. Second row:  Median value of 
$L_{\rm FIR}$, only for detections.
{\it Column 6}:  Total number of galaxies with IRAS data in the subsample.
{\it Column 7}:  Number of upper limits in FIR (at 60 or 100~$\mu$m). 
{\it Column 8}:  First row: Mean value of $R = \log(L_{\rm FIR}/L_{B})$, 
using the Kaplan-Maier estimator from ASURV. Second row: Median value of $R$, 
only for detections.  
\label{tab_average}
\end{table*}

Figure~\ref{lfir_histo} shows the distribution of FIR luminosity for
the optically complete AMIGA sample and in Table~\ref{tab_average} the mean
and median values are given. We include in Fig.~\ref{lfir_histo}  individual 
histograms for: 1)
galaxies detected at both 60 and 100~$\mu$m, 2) those not detected at
one or both wavelengths and 3) the distribution calculated using
survival analysis that takes upper limits into account.  We use the
ASURV package for the latter calculations throughout this 
paper\footnote{Astronomy Survival Analysis (ASURV) Rev. 1.1
is a generalised statistical package that implements the methods presented 
by Feigelson \& Nelson (\cite{feigelson85}) and Isobe et al. (\cite{isobe86}), 
and is described in detail in Isobe \& Feigelson (\cite{isobe90-1}) and 
La Valley et al. (\cite{lavalley92}).}.  
The distribution peaks in the bin
$\log(L_{\rm FIR}/L_{\sun}) = 9.5$--9.75 with the ASURV estimated mean 
$\log(L_{\rm FIR}/L_{\sun}) = 9.15$ (see Table~\ref{tab_average}). 
Practically all galaxies have FIR luminosities between
$\log(L_{\rm FIR}/L_{\sun}) = 7.5$ and $\log(L_{\rm FIR}/L_{\sun}) = 11.25$. 
Only one object, the faint irregular
member of the Local Group CIG~388 ($\equiv$ Sextans B), 
shows $\log(L_{\rm FIR}/L_{\sun})= 6.01$.
It is remarkable that  the bulk of the FIR luminosities (98\%) 
lies below $\log(L_{\rm FIR}/L_{\sun})= 10.5$. 

\begin{figure}
\resizebox{1.0\hsize}{!}{\includegraphics{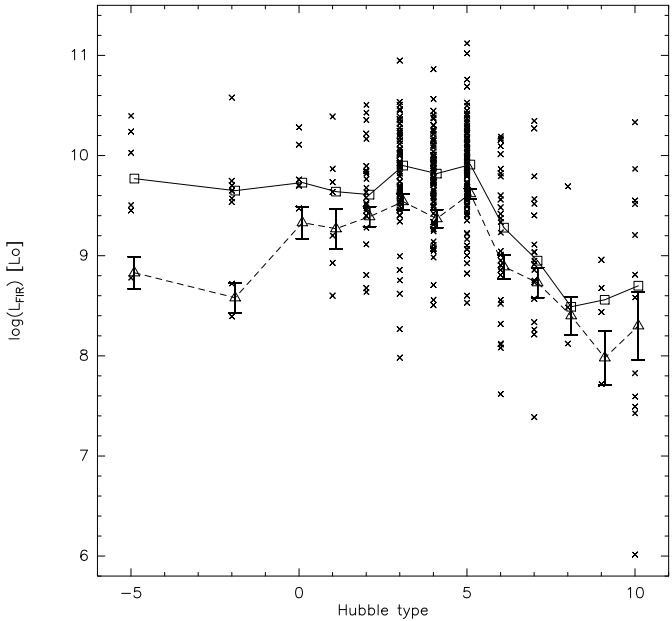}}
\caption{Distribution of FIR luminosity as a function of
Hubble type. Only detections are shown. The open triangles give the mean value
for each Hubble type, calculated with ASURV and taking the upper limits into account.
The open squares are the median values for the detections only.
}
\label{lfir-hubble}
\end{figure}

In Fig.~\ref{lfir-hubble} we show the distribution of the 
FIR luminosities as a function of Hubble type, and in Table~\ref{lfir_histo}
we list the mean (taking upper limits into account with ASURV) and median
(for detections only) values.
The mean values show a strong increase after the early-types
(E--S0) beginning at $T=0$ (S0a) and increasing through $T=5$ (Sc), followed
by a decline to a near constant mean for the latest types ($T=8$--10). ASURV
means for early-types are most strongly driven by upper limits with
most detected E--S0 showing $L_{\rm FIR}$ values above the computed means,
similar to those for
late-type spirals.  This marks the detections as unusual indicating
that these may not be typical (or even) E--S0 galaxies (see discussion
in Paper~II). As we proceed from left to right in the plot the effect 
of upper limits gradually decreases and mean and median values converge. 
Our previously identified (Paper~II)
dominant ($\sim$65\%) isolated late-type ($T=3$--5) spiral population shows
FIR luminosities strongly concentrated (due to the minimisation of
nurture-driven dispersion) in the range 9.4--10.5.  
%
%
We also observe a small but significant population of
spiral types $T= 2$--7 with very low FIR luminosities. We see an apparent
strong drop in mean FIR luminosity ($\sim$0.7 in $\log(L_{\rm FIR})$) later than
type $T=5$.  If real, there are three candidate explanations: 1) decreasing
dust mass simply following decreasing galaxy mass for Scd--Sd, 2)
decreasing dust content in Scd--Sd or 3) less efficient star formation
in Scd--Sd (always relative to Sb--Sc).  The latest types show minimal
upper limits since they are  very local. This mostly dwarf galaxy population 
falls out of our magnitude limited sample beyond a few 
$\times$10$^3$~km\,s$^{-1}$ recession velocity.  

\subsection{FIR luminosity function}

Since the AMIGA sample is optically selected we derive the FIR luminosity 
function (FIRLF) from the optical luminosity function and the fractional 
bivariate function between FIR luminosity and optical luminosity (see Paper~I).
The differential FIRLF, which gives the number density of galaxies per unit 
volume and per unit $\log(L_{\rm FIR})$ interval is derived from the following 
formula:
\begin{equation}
\Psi(L) = 2.5 \Delta M \sum_j \Theta(L|M_j) \Phi(M_j),
\end{equation}
where $L=\log($$L_{\rm FIR}$), and $\Psi$ is the differential FIRLF. 
The bivariate (optical, FIR) luminosity function $\Theta(L|M_j)$ is defined as
\begin{equation}
\Theta(L|M_j) =\frac{P(L,M_j)}{\Delta L},
\end{equation}
where $\Delta L = 0.25$ and $P(L,M_j)$ is the conditional
probability for a source with absolute magnitude $M$ $(M_j+0.5 \Delta
M \ge M > M_j - 0.5 \Delta M)$ to have the logarithm of its FIR
luminosity, $\log(L_{\rm FIR})$, within the interval $[L-0.5 \Delta L,L+0.5
\Delta L]$. The Kaplan-Meier estimator (Schmitt \cite{schmitt85}; Feigelson \&
Nelson \cite{feigelson85}), which also exploits the information content of 
upper limits, has been used in computing the bivariate luminosity function
and the associated errors.  $\Phi$ is the differential OLF per unit
volume and per unit magnitude interval, $\Delta M$ is the bin width
of the OLF in magnitude units.  The factor 2.5 arises because a unit
magnitude interval corresponds to only 0.4 unit of $L$. The summation is
over all bins of the OLF. The errors of $\Psi(L)$ are the quadratic sum
of the uncertainties for the OLF and bivariate luminosity function.

\begin{figure}
\resizebox{1.0\hsize}{!}{\includegraphics[clip=0,width=10.cm,angle=0]{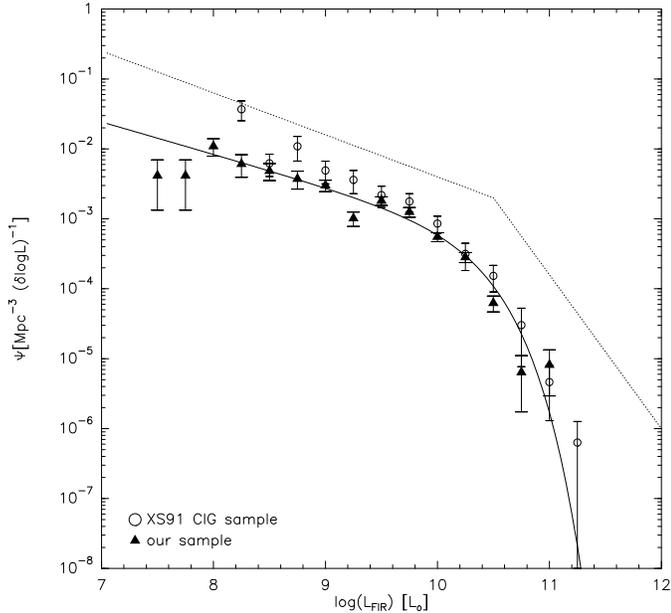}}
\caption{Bivariate FIR luminosity function of our sample compared to the CIG 
sample used in XS91. The dotted line is the fit with a double power-law 
derived in Sanders et al. (\cite{sanders03}) for the Bright IRAS galaxy 
sample. The full line is a Schechter fit to our data.}
\label{lum-function}
\end{figure}

Figure~\ref{lum-function} shows the resultant FIRLF and Table~\ref{lum_function}
lists the corresponding values. Also shown is the
FIRLF from XS91 for a smaller subsample of the CIG. We see that the general
shape has not changed substantially with the use of a larger and more
complete sample. It is our contention that it represents the best
``natural'' or ``nurture-free'' FIRLF yet derived. 
A strong decline in the FIRLF above $\log(L_{\rm FIR}/L_{\sun})\sim 10.5$ is
clearly visible. In the last few bins there are only very few objects 
(see Table~\ref{lum_function}), making the value of the FIRLF uncertain.

 \begin{table}
\caption{FIR luminosity function}
\begin{tabular}{cccc}
\hline
\hline
$\log(L_{\rm FIR})$ (in $L_{\sun}$) & $\Psi$ (Mpc$^{-3} \log(L)^{-1}$) & $n$\\ 
  \hline
       7.50 &     $4.15\times 10^{-3}  \pm  2.82\times 10^{-3}$  & 2.5  \\
       7.75 &     $4.15\times 10^{-3}  \pm  2.82\times 10^{-3}$  & 2.5  \\
       8.00 &     $1.09\times 10^{-2}  \pm  3.05\times 10^{-3}$  & 19.6 \\
       8.25 &     $6.09\times 10^{-3}  \pm  2.17\times 10^{-3}$  & 28.7 \\
       8.50 &     $4.85\times 10^{-3}  \pm  1.32\times 10^{-3}$  & 23.0 \\
       8.75 &     $3.73\times 10^{-3}  \pm  1.07\times 10^{-3}$  & 34.9 \\
       9.00 &     $3.01\times 10^{-3}  \pm  5.63\times 10^{-4}$  & 101.2\\
       9.25 &     $1.02\times 10^{-3}  \pm  2.35\times 10^{-4}$  & 38.1 \\
       9.50 &     $1.81\times 10^{-3}  \pm  2.56\times 10^{-4}$  &123.9 \\
       9.75 &     $1.25\times 10^{-3}  \pm  1.98\times 10^{-4}$  &132.5 \\
       10.00 &    $5.51\times 10^{-4}  \pm  8.06\times 10^{-5}$ & 95.5  \\
       10.25 &    $2.83\times 10^{-4}  \pm  4.60\times 10^{-5}$ & 61.5  \\
       10.50 &    $6.26\times 10^{-5}  \pm  1.59\times 10^{-5}$  & 21.0 \\
       10.75 &    $6.40\times 10^{-6}  \pm  4.66\times 10^{-6}$ & 3.1  \\
       11.00 &    $8.17\times 10^{-6}  \pm  5.23\times 10^{-6}$ & 3.0   \\
\hline
\end{tabular}

The entries are: 
{\it Column 1}: Center of luminosity bin.
{\it Column 2}: Bivariate FIR luminosity function and its error. 
{\it Column 3}: Number of galaxies in the bin. The numbers are not 
integer due to the survival analysis applied.
\label{lum_function}
\end{table}

We have fitted the FIRLF with a Schechter function:

\begin{equation}
\Psi(L) = \Psi_0\left(\frac{L}{L^\star}\right)^\alpha 
\exp\left(-\frac{L}{L^\star}\right).
\end{equation}

The best-fit parameters are $\Psi_0=(7.4\pm1.4)\times 10^{-4}$ Mpc$^{-3}$ 
($\delta \log(L_{\rm FIR}))^{-1}$,
$L^\star =  (1.9\pm0.2) \times 10^{10} L_{\sun}$ and $\alpha = -0.46\pm0.05$.
We have also plotted in Fig.~\ref{lum-function} the fit to the IRAS
Bright galaxy Sample FIRLF (Sanders et al. \cite{sanders03}). They found, in
agreement with other FIR selected samples, that a double-power law
provides the best fit to the data.  The difference from a Schechter fit
typically starts to be noticeable above $10^{11}L_{\sun}$.  Sulentic \& Rabaca 
(\cite{sulentic94}) earlier pointed out the difficulty with using a Schechter
function to adequately describe nurture-affected samples. With only
three galaxies above $\log(L_{\rm FIR}/L_{\sun})=11.0$ our sample is well fit 
by a Schechter function.
 
\subsection{$L_{\rm FIR}$\ and $L_{B}$}
\label{lfir_and_lb}

\begin{figure}
\resizebox{1.0\hsize}{!}{\includegraphics{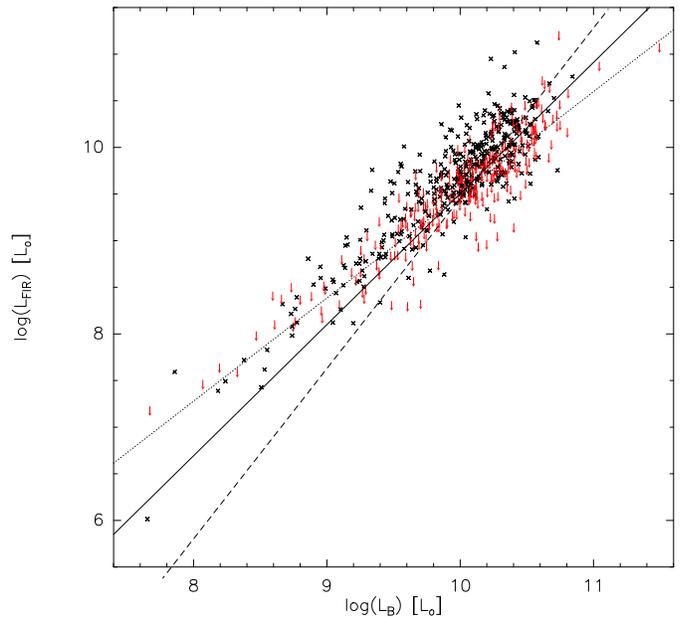}}
\caption{
$L_{\rm FIR}$\ vs. $L_{B}$\ for the optically complete, magnitude limited 
subsample  ($n=701$, see Sect.~\ref{luminosity} for exact definition).
The full line indicates the best-fit bisector slope derived with ASURV, 
the dotted line shows the result of the regression adopting $L_{B}$, and 
the dashed line adopting $L_{\rm FIR}$\ (dashed) as independent variable.
}
\label{lfir_lb_all}
\end{figure}

Figure~\ref{lfir_lb_all} plots  $L_{\rm FIR}$ vs.  $L_{B}$ for the optically
complete sample defined in Sect.~\ref{sample_definition}. 
$L_{B}$ was calculated as $L_{B} = 10^{(1.95-0.4M_{\rm zw-corr})}$ 
in units of solar bolometric luminosity
where $M_{\rm zw-corr}$ is the absolute Zwicky magnitude corrected for
systematic errors, Galactic extinction, internal extinction and with
K-correction applied (see Sect.~\ref{sample_definition} and Paper~I).
This definition\footnote {Note that this definition differs by a
factor of 1.7 from the definition used in Paper~I which was normalised
to the solar luminosity in the blue.} 
provides an estimate of the blue luminosity ($\nu
L_\nu$) at  4400 \AA. 
In Fig.~\ref{lfir_lb_all} we see scatter due
to measurement uncertainties and intrinsic dispersion. The latter
should be minimised as much as possible to nature-driven dispersion.
In this sample we have reduced dispersion due to both one-on-one
interactions and environmental density.  
Most galaxies lie close to the correlation with a dispersion of 0.23
for the detected galaxies. There are  four clear outliers close to
$L_{\rm FIR}$$=10^{11} L_{\sun}$. Three of them have been classified as 
possibly interacting in Paper~II.

We applied linear regression analysis in order to estimate the
functional relationship between the two variables.  
Since we are interested in the physical relation between the two
variables and since both variables have intrinsic uncertainties we
decided to use a symmetric method. We derived the regression
coefficients for both $L_{\rm FIR}$ vs. $L_{B}$  and $L_{B}$ vs. 
$L_{\rm FIR}$ using ASURV and
calculated the bisector regression line shown in Fig.~\ref{lfir_lb_all} 
from these, following the formula in
Isobe et al. (\cite{isobe90-2}). We used the Schmitt's binning method 
(Schmitt \cite{schmitt85})
as the only method offered by ASURV able to deal with censored data in
the independent variable.  We note however, that for the cases where
the other two methods (estimation-maximisation method, and
Buckley-James method) could be applied, a satisfactory agreement was
found. The results for the linear regression (ASURV bisector) are
listed in Table~\ref{regression}. 
\begin{table*}
      \caption{Correlation analysis of $L_{\rm FIR}$\ vs. $L_{B}$.}
\begin{tabular}{lcccccc}
\hline
\hline
(1) & (2) & (3) & (4) & (5) & (6) & (7)  \\
Sample & $n$ & $n_{\rm up}$ & $slope$ & $intercept$ & $slope$ & $intercept$ \\
       &   &              & (bisector) & (bisector) & ($L_{B}$ indep.) & ($L_{B}$ indep.) \\
\hline
Total  &701 & 312& 1.41$\pm$0.02  & $-$4.55$\pm$0.25 &  1.11$\pm$0.03 & $-$1.57$\pm$0.34 \\
\hline
Sa--Sab (1--2)  &  49& 18 & 1.37$\pm$0.09  & $-$4.29$\pm$0.87 & 0.87$\pm$0.23 & 0.76$\pm$1.67 \\
Sb--Sc (3--5)  & 452&181 & 1.35$\pm$0.03  &  $-$3.98$\pm$0.31 & 1.04$\pm$0.06 & $-$0.77$\pm$0.57\\
Scd--Im (6--10) & 115& 49 & 1.25$\pm$0.03 & $-$2.99$\pm$0.30 & 1.16$\pm$0.05 & $-$2.09$\pm$0.58\\
\hline
Interacting      & 14 &  2 & 1.52$\pm$0.12  & $-$5.25$\pm$1.15 & 1.43$\pm$0.13 & $-$4.34$\pm$1.32 \\
%
\hline
\label{regression}
\end{tabular}

The slope and intercept are defined as: 
$\log(L_{\rm FIR}) = \log(L_{B})\times slope + intercept$. The entries are: 
{\it Column 1}: Subsamples considered. 
All subsamples are selected from the optically complete, magnitude limited subsample
(see Sect.~\ref{sample_definition}). 
In the early-type subsamples (E and S0)  the relative number of undetected galaxies in $L_{\rm FIR}$\ 
is very high so that a regression slope could not be determined. 
{\it Column 2}: Total number of galaxies in the respective samples. 
{\it Column 3}: Number of galaxies with upper limits in FIR. 
{\it Column 4}: Bisector slope and its error of the best-fit regression line
derived with the Schmitt binning method in the ASURV package. 
{\it Column 5}: Bisector intercept. 
{\it Column 6}: Slope and its error of the best-fit regression line
derived with the Schmitt binning method in the ASURV package adopting $L_{B}$\ as
independent variable. 
{\it Column 7}: Bisector intercept adopting $L_{B}$\ as
independent variable. 
\label{tab_regression}
\end{table*}
The alternate approach would be to
compute the regression assuming that optical luminosity is the
independent variable. The results are also listed in Table~\ref{regression}
and show that the conclusions drawn in the following would not be 
substantially changed if $L_{B}$ had been adopted as independent variable. 
The best-fit slope for the entire sample is
$L_{\rm FIR} \propto L_{B}^{1.41\pm0.02}$.  Our slope is 
shallower than the one found by Perea et al. (\cite{perea97}) for a smaller
subsample of the CIG, $L_{B} \propto L_{\rm FIR}^{0.65\pm0.09}$ 
(giving a slope of the
inverse relation of $L_{\rm FIR}\propto L_{B}^{1.54}$).  
The main reason for this difference is our use of the bisector slope, 
whereas Perea et al. (\cite{perea97}) derived the
slope with $L_{\rm FIR}$\ as independent variable. With our larger sample we
derive a similar slope when adopting  $L_{\rm FIR}$ as independent variable
($L_{B}\propto L_{\rm FIR}^{0.55\pm0.03}$).  For the present
data set, however, we think that the bisector slope (or $L_{B}$ as
independent variable) is the better choice for investigating the
functional relation between both variables. A possible explanation for the 
slope $>$1, suggested by Perea et al. (\cite{perea97}), 
is an increase of the dust extinction with galaxy luminosity,
yielding a faster increase of the FIR emission in comparison
to the extinction-affected blue luminosity. An alternative reason could be 
the increase of the recent star formation (SF) activity 
(traced by $L_{\rm FIR}$) with galaxy luminosity.

\begin{figure*}
\resizebox{0.5\hsize}{!}{\includegraphics{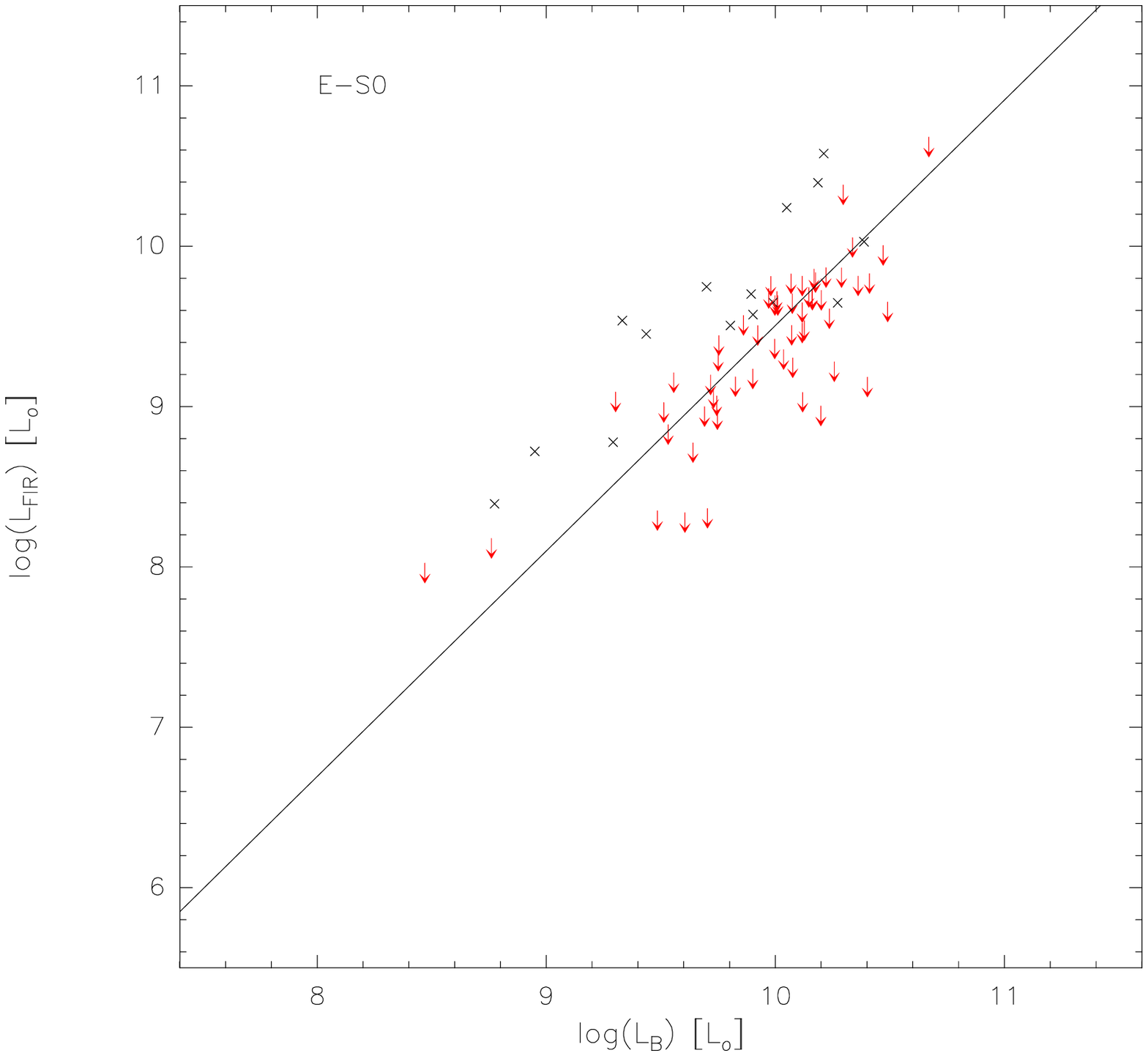}}
\resizebox{0.5\hsize}{!}{\includegraphics{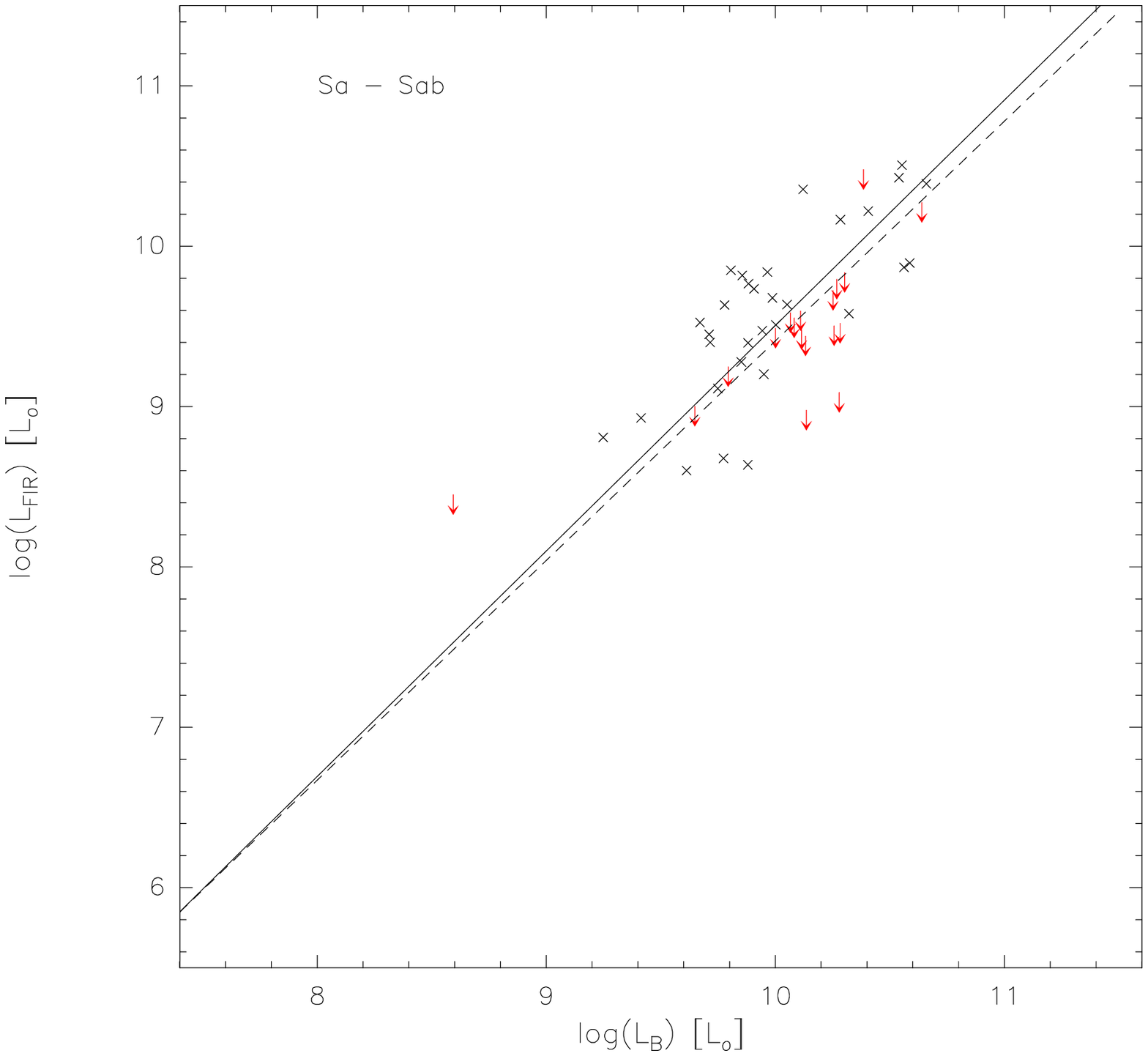}}
\resizebox{0.5\hsize}{!}{\includegraphics{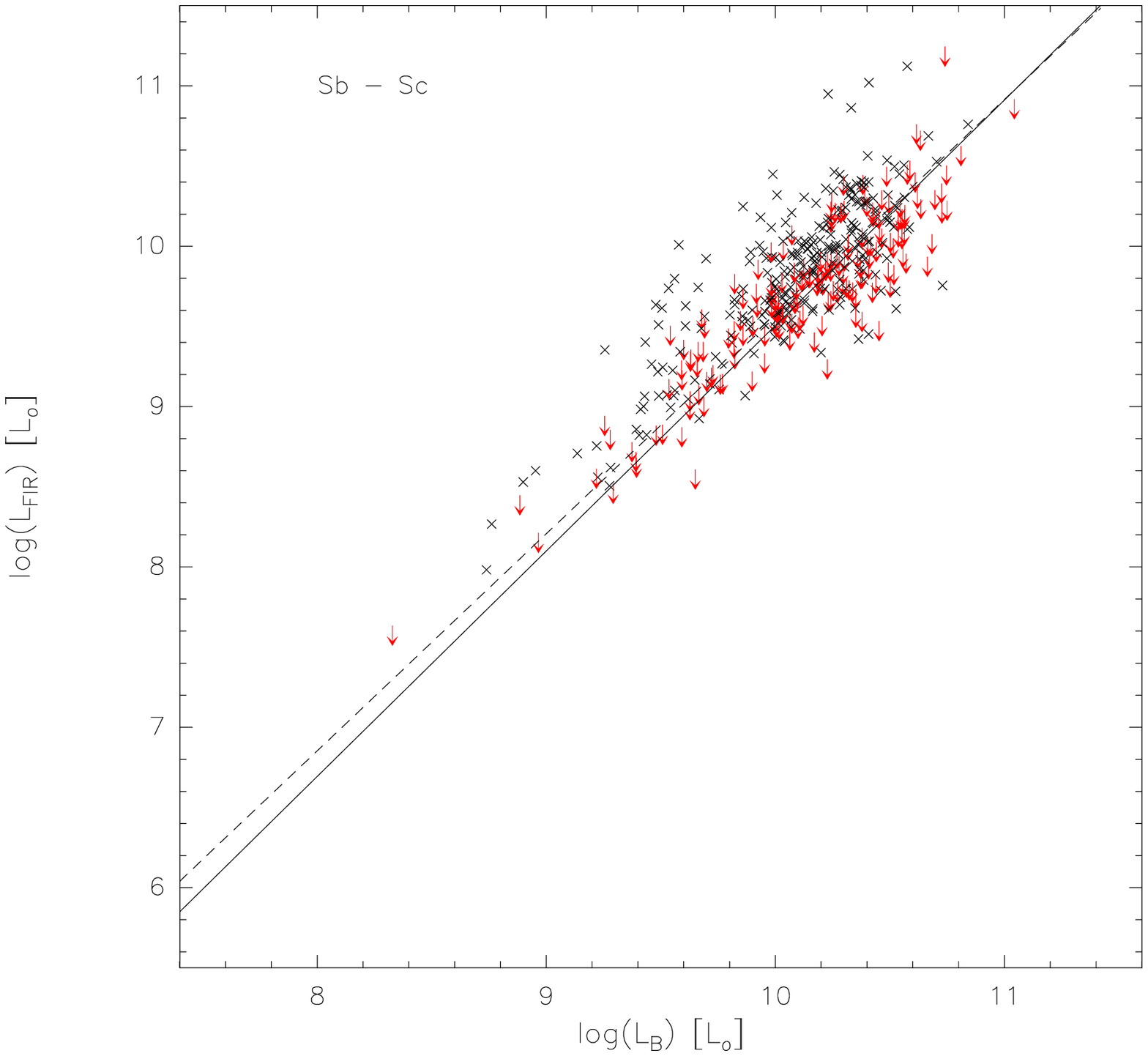}}
\resizebox{0.5\hsize}{!}{\includegraphics{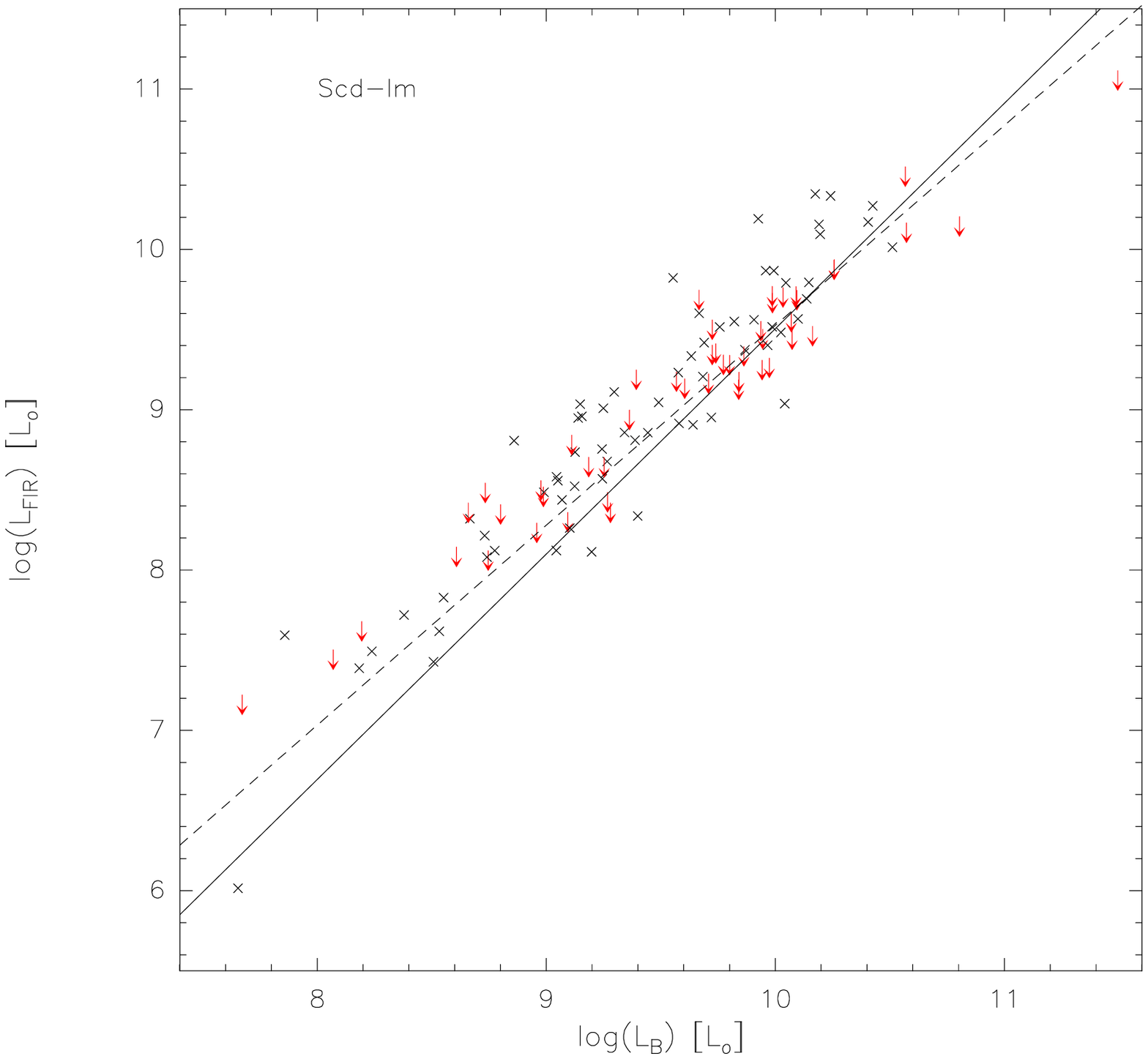}}
\caption{$L_{\rm FIR}$\ versus $L_{B}$\ for subsamples of different Hubble 
types. The full line is the bisector fit for the total AMIGA sample
presented in Fig.~\ref{lfir_lb_all}, whereas the dashed line is the bisector 
fit for the present subsample. For the early type subsample (E and E--S0a) 
no reliable regression fit could be
derived due to the large number of upper limits.
}
\label{lfir_lb_reg_hubble}
\end{figure*}

Figure~\ref{lfir_lb_reg_hubble} presents $L_{\rm FIR}$ vs. $L_{B}$ for 
subsamples of different Hubble types. Due to the low detection rate for 
early-type
galaxies (E--S0), no reliable regression fit could be derived for this
subsample. The correlation for the early-types shows evidence for a composite
population with typical FIR deficient galaxies below the superimposed
regression line and overluminous galaxies, showing a roughly linear correlation
spanning 2 dex, above the line. As mentioned before, IR overluminous
early-type galaxies must be regarded with suspicion until their
morphologies are confirmed with higher resolution images than the POSS2
used for our morphology revision.  At the same time, {\it bona fide}
isolated early-types are of particular interest in view of ideas that
see all of them as a product of nurture (mergers/stripping/harassment).  
There are only small differences in the measured slopes
(see Table~\ref{regression} and Fig.~ \ref{lfir_lb_reg_hubble}) 
of least-squares regression  lines as a
function of Hubble type.
%

\begin{figure}
\resizebox{1.0\hsize}{!}{\includegraphics{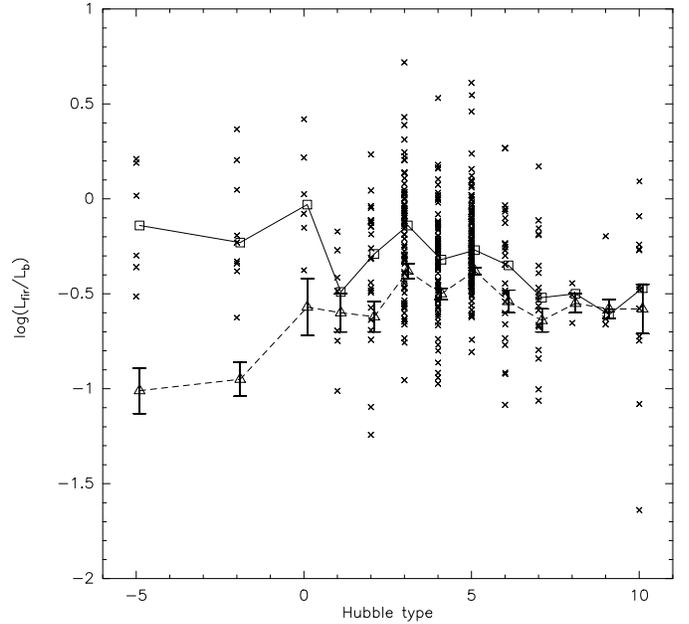}}
\caption{$R= \log(L_{\rm FIR}/L_{B}$), for the optically complete sample as
a function of Hubble type. Only detections are shown. The open triangles give 
the mean value for each Hubble type, calculated with ASURV, taking the upper 
limits into account. The open squares are the median values for the detections.
}
\label{r_hubble}
\end{figure}

Finally, we derive the distribution and the mean value of  
$R=\log(L_{\rm FIR}/L_{B})$, a variable frequently used as an indicator of 
SF activity. In Table~\ref{tab_average} we list
the average and median values of $R$, together with those of 
$L_{\rm FIR}$ and $L_{B}$ derived for different subsamples.
Figure~\ref{r_hubble} shows $R$ as a function of the morphological
subtypes.  No clear trend of $<R>$ is found within the spiral
galaxies with $<R>$ essentially constant between $T=1$--7 (Sa--Sd).
$<R>$, as well as $<L_{\rm FIR}>$ (Fig.~\ref{lfir-hubble}),
is significantly lower for early-types (E and S0) 
although values derived using survival analysis might be uncertain due
to the large number of upper limits. This means that early-type
galaxies have a lack in FIR emission with respect to their blue
luminosity with the ones showing values similar to spirals possibly  being
misclassified spirals. Late type galaxies (Sd--Im) are on average less
luminous both in $L_{\rm FIR}$ and $L_{B}$, but show the same  $<R>$ as spirals.

\subsection{IRAS colours}

IRAS flux ratios provide another potentially useful diagnostic.
$F_{\rm 60}/F_{\rm 100}$ (Telesco et al. \cite{telesco88}), 
$F_{\rm 25}/F_{\rm 60}$ (XS91) and $F_{\rm 12}/F_{\rm 25}$ 
(Bushouse et al. \cite{bushouse88}) 
have been used as environmental diagnostics. 
For example, $F_{\rm 60}/F_{\rm 100}$ measures the 
dust temperature and has been found  to  increase with  the level
of star formation activity (de Jong et al. \cite{dejong84}). 
$F_{\rm 25}/F_{\rm 60}$ is an indicator for AGN activity with values above  
$F_{\rm 25}/F_{\rm 60} = 0.3$ regarded as indicative of a Seyfert nucleus 
(de Grijp et al. \cite{degrijp85}).

\begin{figure}
\resizebox{0.8\hsize}{!}{\includegraphics{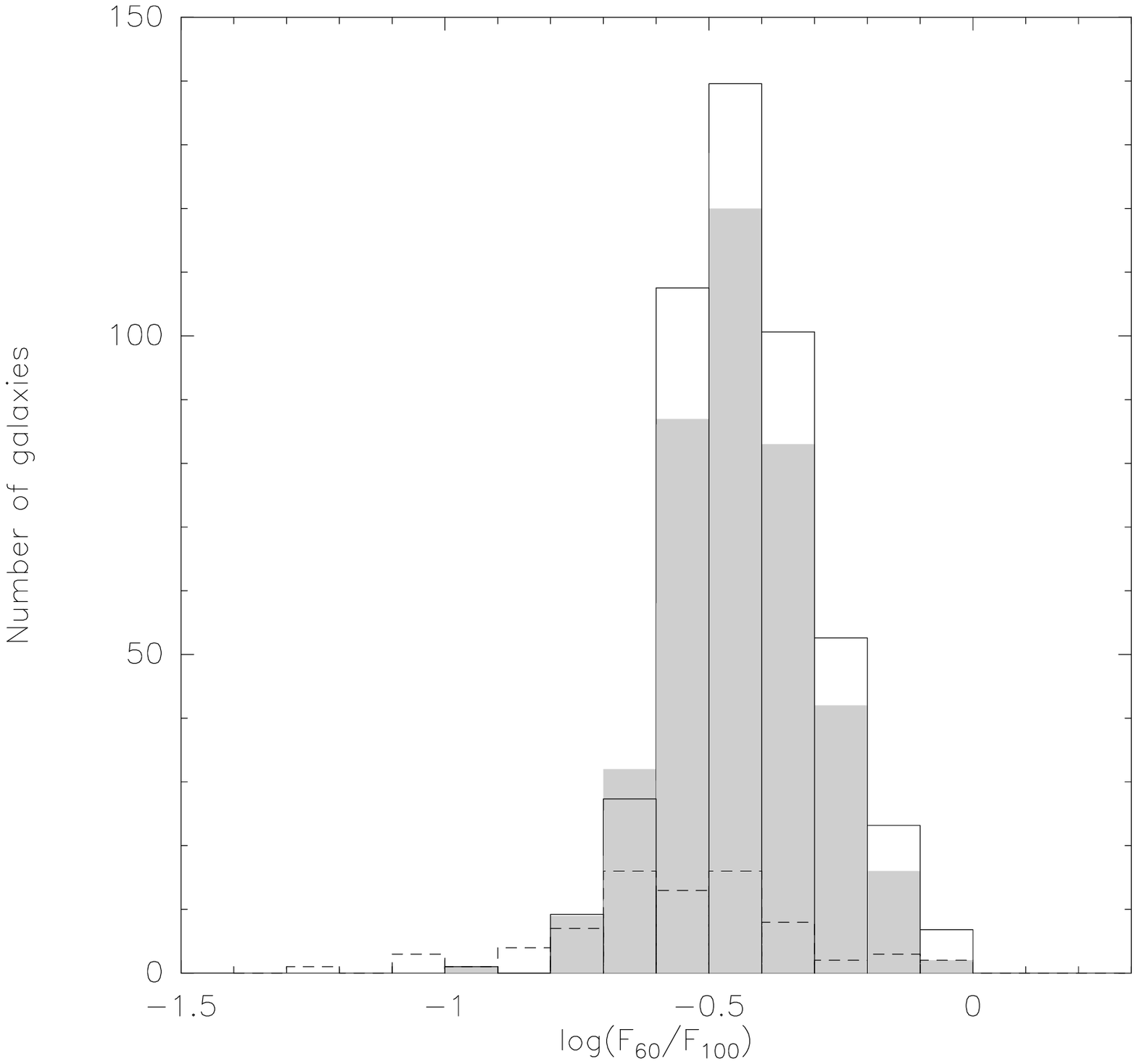}}
\\
\resizebox{0.8\hsize}{!}{\includegraphics{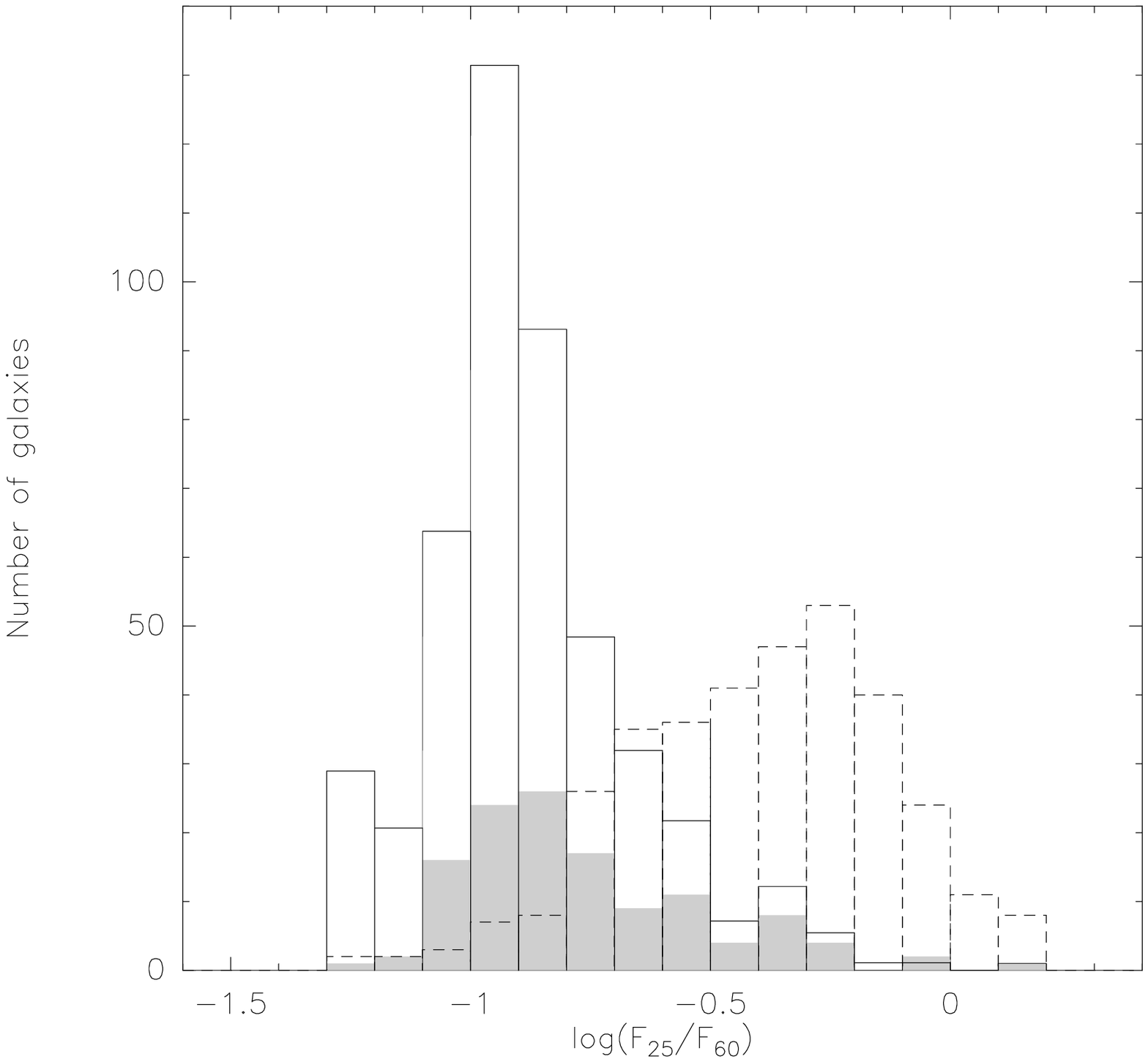}}
\\
\resizebox{0.8\hsize}{!}{\includegraphics{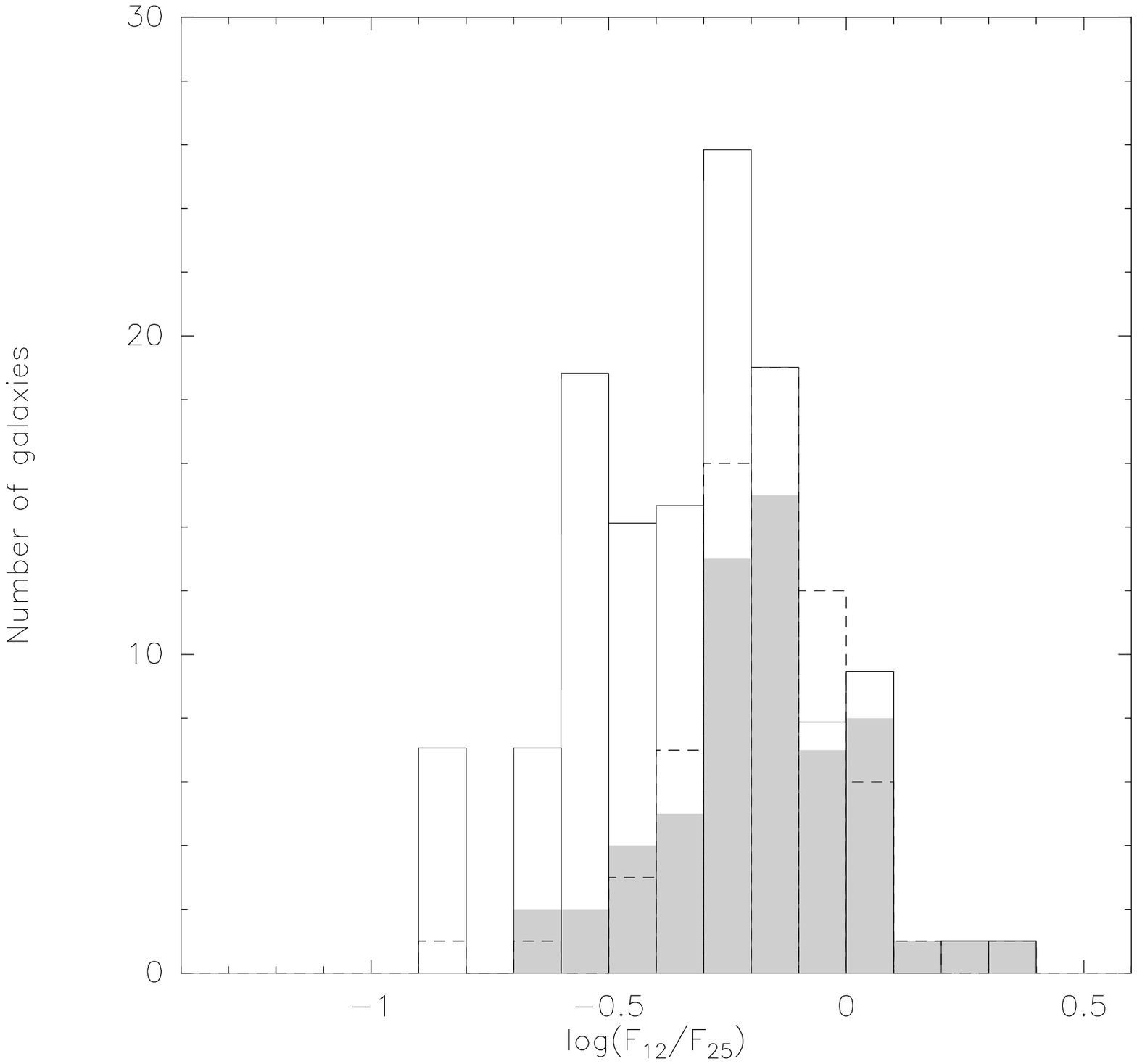}}
\caption{
IRAS Colours  for the AMIGA sample. The full line shows the 
histogram calculated with ASURV taking into account censored data points, 
the grey-shaded area shows
the detections and the dashed line the upper limits. 
Only galaxies with detections at 60~$\mu$m are considered for 
$\log(F_{\rm 60}/F_{\rm 100})$ and $\log(F_{\rm 25}/F_{\rm 60})$
, and only galaxies with  detections
at 25~$\mu$m for $\log(F_{\rm 12}/F_{\rm 25})$.}
\label{colors}
\end{figure}

\begin{table*}
      \caption{Mean and median values of IRAS colours.}
\begin{tabular}{lcccccccc}
\hline
\hline
(1) & (2) & (3) & (4) & (5) & (6) & (7) & (8) & (9)  \\
Sample & $n_{\rm tot}$ & $n_{\rm up}$ &$<\log(F_{60}/F_{100})>$ & $n_{\rm up}$ 
& $<\log(F_{25}/F_{60})>$ & $n_{\rm tot}$ 
& $n_{\rm up}$ & $<\log(F_{12}/F_{25})>$\\
 & & & med($\log(F_{60}/F_{100})$) & & med($\log(F_{25}/F_{60})$) 
& & & med($\log(F_{12}/F_{25})$) \\
\hline
\parskip=0pt
Total  &468 & 76& $-$0.42$\pm$0.01 & 343 & $-$0.87$\pm$ 0.02 & 126 & 67 & $-$0.33$\pm$0.03\\
               &    &   & $-$0.45         &     & $-$0.83           &     &    & $-$0.18 \\
S/Im ($T=1$--10) & 443 & 72 & $-$0.43$\pm$0.01 & 326 & $-$0.87$\pm$0.02 & 118 & 65 & $-$0.33$\pm$0.03 \\ 
               &    &       & $-$0.45          &     &  $-$0.83 & & & $-$0.16\\
\hline
E ($-$5) & 9  & 3  &  $-$0.23$\pm$0.06 & 4  & $-$0.73$\pm$0.09 & 5 & 2 & $-$0.47$\pm$0.07 \\
        &     &     & $-$0.23         &      & $-$0.74          &   &  & $-$0.48 \\
S0 ($-$2) & 10 & 1  &  $-$0.39$\pm$0.06 & 8 & $-$1.02$\pm$0.004 & 2 & 0 & $-$0.16$\pm$0.01  \\ 
        &     &     & $-$0.35          &     & $-$1.02                     & &  & $-$0.16 \\ 
S0a (0) &   6 &   0 & $-$0.36$\pm$0.07 & 5   & $-$0.98$^*$ & 1 & 0 & $-$0.27$^*$\\
        &     &     & $-$0.27          &     & $-$0.98          & &  & $-$0.27\\
Sa  (1) &  9 &  2  & $-$0.43$\pm$0.04 & 7   & $-$0.79$\pm$0.05 & 2 & 1 & $-$0.32$^*$\\
        &     &     &$-$0.42           &     & 0.71             &  &   & $-$0.32 \\
Sab (2) & 27  & 3   & $-$0.42$\pm$0.03 & 18  & $-$0.81$\pm$0.04 & 9& 4 & $-$0.39$\pm$0.09\\
        &     &     &$-$0.45           &     & $-$0.72          & &  & $-$0.22\\
Sb  (3) &  88 & 15  & $-$0.41$\pm$0.02 & 63  & $-$0.87$\pm$0.03 & 25 & 17 &$-$0.50$\pm$0.10 \\
        &     &     &$-$0.44           &     & $-$0.89          & &  & $-$0.18 \\
Sbc (4) & 104 & 17  & $-$0.45$\pm$0.01 & 77  & $-$0.83$\pm$0.03 &27  & 15 & $-$0.27$\pm$0.04 \\
        &     &     &$-$0.46           &     & $-$0.78          &    &    & $-$0.15 \\
Sc  (5) & 138 &  24 & $-$0.46$\pm$0.01 & 107 & $-$0.87$\pm$0.03 &32& 14& $-$0.24$\pm$0.05 \\
        &     &     &$-$0.48           &     & $-$0.88          & &  & $-$0.10 \\
Scd (6) &  34 &  6  & $-$0.45$\pm$0.02 & 24  & $-$0.87$\pm$0.06 &10& 6 & $-$0.36$\pm$0.08\\
        &     &     &$-$0.44           &     & $-$0.82          &  & & $-$0.17\\
Sd  (7) &  21 &  2  & $-$0.39$\pm$0.03 & 13  & $-$0.85$\pm$0.07 &8 & 5 & $-$0.23$\pm$0.003 \\
        &     &     &$-$0.40           &     & $-$0.79         & &  & $-$0.23 \\
Sdm  (8)&   4 & 1   & $-$0.40$\pm$0.02 & 3   & $-$0.70$\pm$0.15 &1 & 1 & \_\\
        &     &     & $-$0.40            &     & $-$0.34             & &  & \_ \\
Sm  (9) &  5  & 1   & $-$0.27$\pm$0.02 & 3   & $-$1.19$\pm$0.08 &2 & 1 & $-$0.28$\pm$0.04\\
        &     &     &$-$0.28           &     & $-$1.24             & &  & $-$0.22 \\
Im (10) &  13 &  1  & $-$0.31$\pm$0.03 & 11   & $-$0.95$\pm$0.08 &2 & 1 & $-$0.09$\pm$0.10 \\
        &     &     & $-$0.35          &     & $-$0.80          & &  & 0.06 \\
\hline
Interacting      & 14 &  2 & $-$0.36$\pm$0.03 & 10 &  $-$0.87$\pm$ 0.03 & 4  & 0 & $-$0.32$\pm$0.08\\
        &     &     &$-$0.39           &     &   $-$0.89        & &  &$-$0.34 \\
\hline
\end{tabular}

For the entries marked with ``$^*$'' ASURV was not able to calculate an error.
A ``\_'' means that the entry could not be calculated due to the low number 
of detections.  For ratios involving $F_{\rm 60}$, 
only galaxies with detections at 60~$\mu$m are taken into account and for
$\log(F_{\rm 12}/F_{\rm 25})$ only
 galaxies with detections at 25~$\mu$m.
The entries are:
{\it Column 1}: Considered subsample.
All subsamples are selected from the optically complete, magnitude limited 
subsample. The interacting subsample consists of galaxies 
excluded from the CIG  in Paper~II (see Sect.~\ref{sample_definition}).
{\it Column 2 and 7}: Total number of galaxies in the subsample.
{\it Column 3, 5, and 8}: Number of galaxies with upper limits. 
{\it Column 4, 6 and 9}: First row: Mean value of the ratio, using the Kaplan-Maier
estimator from ASURV.
Second row:
Median value of the same ratio, only for detections.
\label{tab_colors}
\end{table*}

Figure~\ref{colors} presents histograms of different IRAS colours 
for our optically complete subsample. The average and median values  
are listed in Table \ref{tab_colors}. 
The flux ratios log($F_{\rm 60}/F_{\rm 100}$) and log($F_{\rm 12}/F_{\rm 25}$)  
show a relatively symmetric distribution around
the peak values. 
On the other hand, log($F_{\rm 25}/F_{\rm 60}$) exhibits a tail towards high values. 
The relative intensity of this tail weakens when only including detections with
a higher $S/N$ (we used   $S/N>7$ as a test), suggesting that part of it
might be due to uncertain values, mainly at 25~$\mu$m.
Another possible reason for high values of $F_{\rm 25}/F_{\rm 60}$ can be the
presence of AGNs, following the finding of de Grijp et al. (\cite{degrijp85}) 
that galaxies with $F_{\rm 25}/F_{\rm 60}>0.3$ are very likely
to host an AGN.  We have checked the values of $F_{\rm 25}/F_{\rm 60}$ 
for galaxies with an AGN  listed in Sabater et al. (in prep.).
Their list includes galaxies catalogued to have an AGN in NED or in the 
V\'eron-Cetty Catalogue of Quasars and Active Nuclei (V\'eron-Cetty \& 
V\'eron \cite{veron03}), as well as radio-excess
galaxies with radio luminosities more than 5 times the values predicted by the
radio-FIR correlation and which are likely to be radio-loud quasars 
(Sopp \& Alexander \cite{sopp91}). We found that 10 out of 11 active galaxies 
with detections at both 25~$\mu$m and 60~$\mu$m have values
of $\log(F_{\rm 25}/F_{\rm 60})\ge -0.7$, the value where the departure from
symmetry in the distribution of   $F_{\rm 25}/F_{\rm 60}$ starts to be 
noticeable.
Furthermore, 10 out of 14 galaxies with upper limits at  $F_{\rm 25}$  might
lie above this threshold, but the upper limit at  $F_{\rm 25}$ makes a firm
conclusion impossible.
Thus, even though the absolute number of galaxies with known AGNs is
not enough to explain the tail towards high  $F_{\rm 25}/F_{\rm 60}$,
they might be responsible for part of it. 

\begin{figure}
\resizebox{0.8\hsize}{!}{\includegraphics{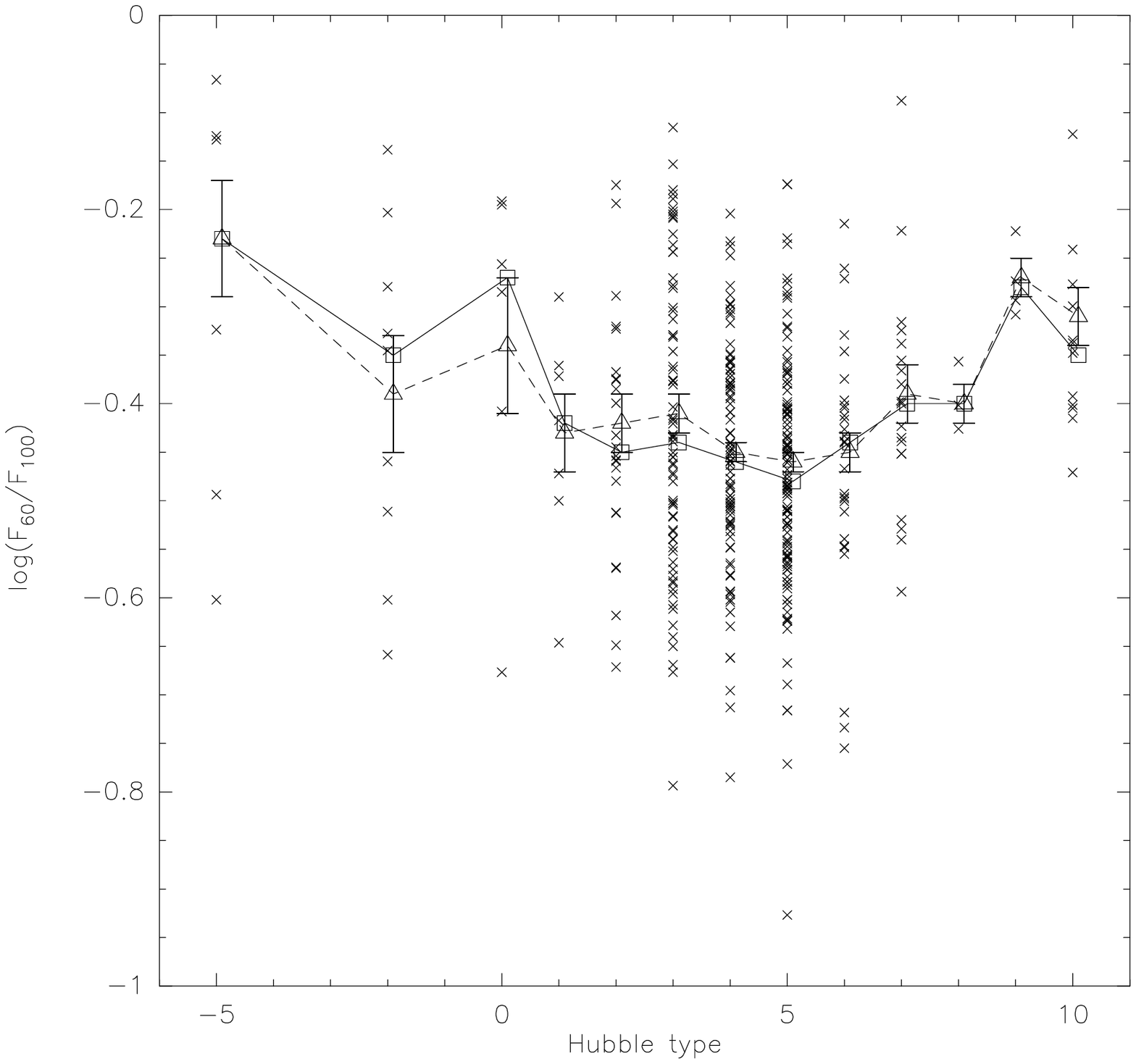}}
\\
\resizebox{0.8\hsize}{!}{\includegraphics{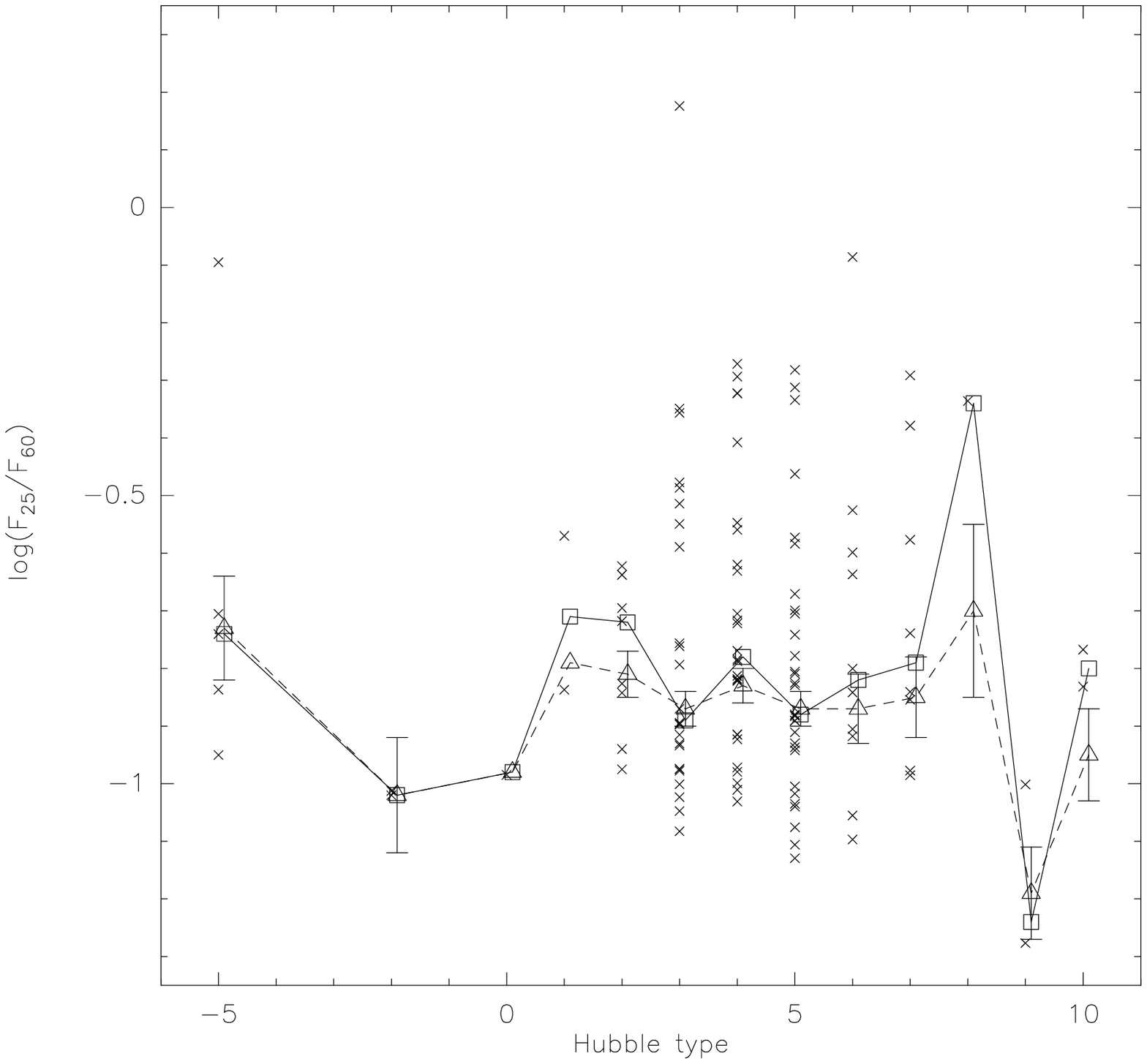}}
\\
\resizebox{0.8\hsize}{!}{\includegraphics{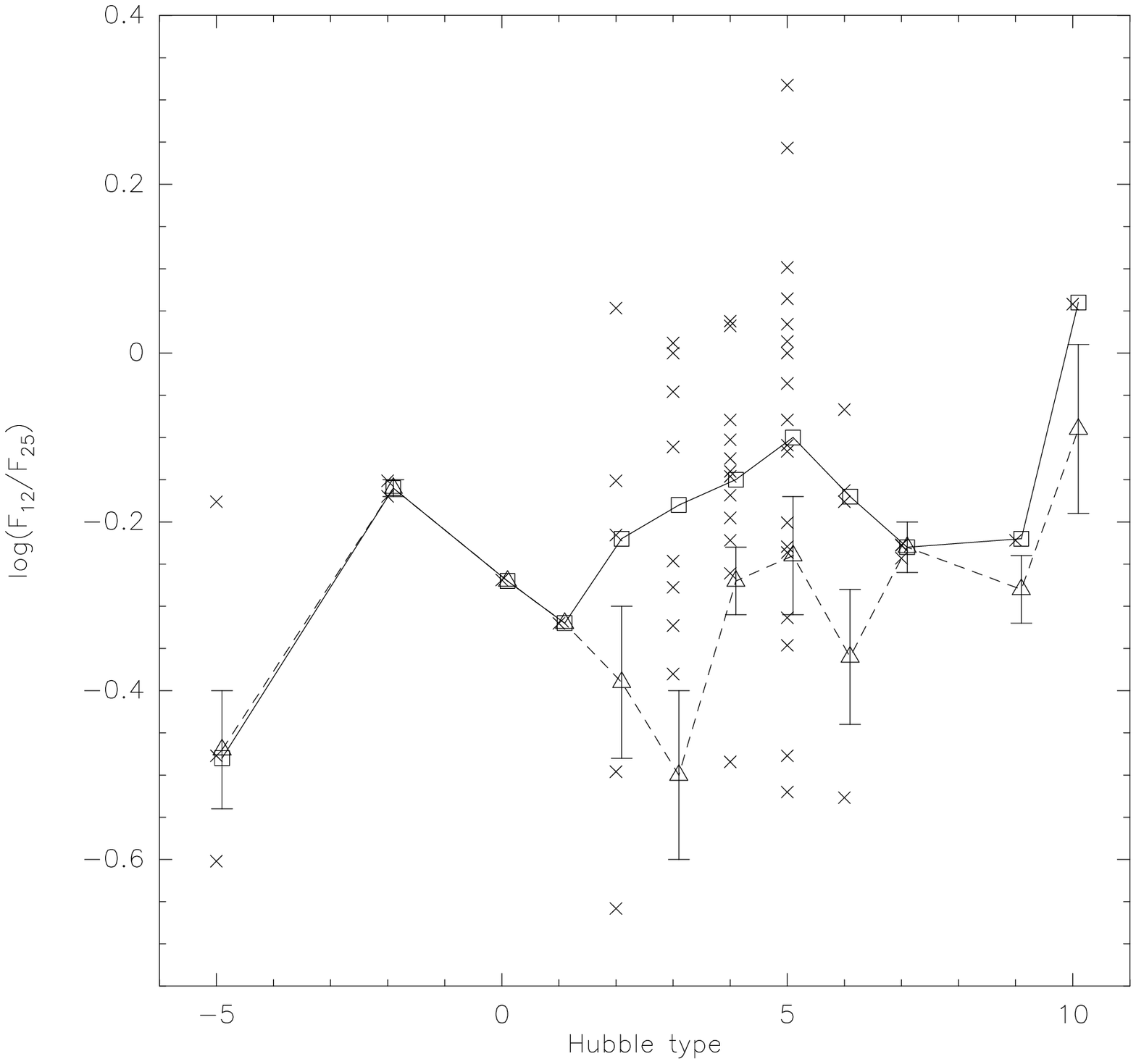}}
\caption{IRAS colours as a function of Hubble type for the optically 
complete sample. For $\log(F_{\rm 60}/F_{\rm 100})$ and  
$\log(F_{\rm 25}/F_{\rm 60})$ 
only galaxies with detections at 60~$\mu$m are taken into account, and for  
$\log(F_{\rm 12}/F_{\rm 25})$ only galaxies with detections at 25~$\mu$m.
Only detected galaxies are plotted (crosses). The open squares indicate the
mean values from Table~\ref{tab_colors}, calulated with ASURV and taking into
account censored data points. The open triangles indicate the
median values based on detections only. When no error bar is given, 
this could not be calculated due to the low number of detections.}
\label{color-hubble}
\end{figure}

In Fig.~\ref{color-hubble} we show the different IRAS colours as a function of 
Hubble type and in Table~\ref{tab_colors} we list the average and mean values.
We notice the following:
\begin{itemize}
\item The range in  $\log(F_{\rm 60}/F_{\rm 100})$
occupied by most galaxies is quite narrow, with almost all  objects 
having $-0.7 \la \log(F_{\rm 60}/F_{\rm 100}) \la -0.2$.
\item For $\log(F_{\rm 60}/F_{\rm 100})$, we find a clear trend with Hubble 
type. The value is highest for ellipticals (type $-$5), decreasing towards 
spirals and increasing again for late-type galaxies, starting from type 7--8,
until irregulars (type 10).
\item There is no significant trend in Hubble type visible for 
$\log(F_{\rm 25}/F_{\rm 60})$ or
$\log(F_{\rm 12}/F_{\rm 25})$. The low number of detections
might be the reason. We neither found  a trend for
$\log(F_{\rm 12}/F_{\rm 60})$ (not shown here), for which we derived a mean value
for the total sample of $-$1.13$\pm$0.02, and very similar values
for each Hubble type individually. 
\end{itemize}


\section{Discussion\label{section5}}

\subsection{Comparison to other non-interacting samples}
\label{comparision_non_interacting}

\subsubsection{$L_{\rm FIR}$\ and $L_{B}$}
\label{lfir_lb_comparison}

We compare the distribution of the FIR luminosity and of $R$ to 
that of the galaxy sample of the Center of Astrophysics (CfA, Huchra et 
al.~\cite{huchra83}), whose FIR properties, based on data of the IRAS FSC,  
were studied in Thuan \& Sauvage (\cite{thuan92}) (hereafter TS92) and 
Sauvage \& Thuan (\cite{sauvage92}) (hereafter ST92). The
CfA sample consists of 2445 galaxies representing a complete flux-limited
sample ($m_{\rm zw} \le 14.5$) selected in Galactic coordinates.
No selection with respect to environment was carried out.
In order to properly compare the two data sets we
applied  the same magnitude cutoff as in TS92 ($m_{\rm zw} \le 14.5$, 
in {\it uncorrected} Zwicky magnitudes), to our sample. 
We then compared the velocity distribution of these two samples 
(the CfA sample and our adapted sample) and found a very
good agreement, with only two differences:
in the CfA sample the  peak at $\sim$5000~km\,s$^{-1}$ is missing
due to their restriction in coordinates which avoids the
region of the Perseus-Pisces supercluster responsible for this peak.
Furthermore, in our sample  with the above magnitude restriction there were no
galaxies beyond 8500~km\,s$^{-1}$, 
whereas about 4\% of the galaxies in the CfA sample
have velocities above this value. We checked the effect of excluding these
high velocity galaxies in the CfA sample on the subsequent results and 
did not notice any 
significant differences. 

In order to correctly compare the luminosity distributions,
we derived the distances for the CfA sample in the same way
as for our galaxies: for close-by galaxies 
($V_{\rm hel}<1000$~km\,s$^{-1}$) (for which
we used redshift-independent distances from the literature) 
we adopted the distances given by TS91, who used
a Virgo-infall model to calculate them. 
For galaxies with  $V_{\rm hel}>1000$~km\,s$^{-1}$ we calculated
the velocities after the 3K correction, $V_{\rm 3K}$, in the same way 
as for the AMIGA sample (see Paper~I), and derived the
distances as $D = V_{\rm 3K}/H_0$. We  used
the same Hubble constant  ($H_0 = 75$~km\,s$^{-1}$\,Mpc$^{-1}$)
in both samples.

As a test to find possible systematic differences we
compared the distances,  
$L_{\rm FIR}$, $L_{B}$ and $R$ for those galaxies that are common in both 
samples (total: $n = 98$, with FIR detections in both samples: $n = 87$). 
TS92 used $B_T^0$ to derive $L_{B}$. For the CfA sample we
estimated the corrected Zwicky magnitudes from  $B_T^0$ using 
the linear relation found between both quantities in paper I.
Then we calculated $L_{B}$ with the same formula as for the AMIGA sample.
The calculation of $L_{\rm FIR}$ was also done in the same way for both
samples.
For the 98 galaxies we found an excellent correlation between
the distances used by us and those based on data of TS92, with a correlation
coefficient of 0.995 and a slope of $1.01\pm 0.01$.
We also found a very good correlation
between our values of $L_{\rm FIR}$ and the values derived by TS92
(correlation coefficient of 0.96 for detections) 
as well as for $L_{B}$\ (correlation coefficient of 0.90) 
and for $R$ (correlation coefficient of 0.85 for detections). 
The mean values of $\log(L_{\rm FIR})$, $\log(L_{B})$ and $R$ for the 
galaxies 
in common practically agree (see Table~\ref{comp_sauvage}), showing that a
comparison of both data sets is justified. 

\begin{table*}
\caption{Comparison to the CfA sample (Thuan \& Sauvage \cite{thuan92}).}
\begin{tabular}{lcccccccc}
\hline
\hline
(1) & (2) & (3) & (4) & (5) & (6) & (7) & (8) & (9) \\ 
Subsample  & $n$ &  $<\log(L_{B})>$ & $\sigma_{B}$ & $n_{\rm up}$ & 
$<\log(L_{\rm FIR})>$ & $\sigma_{\rm FIR}$ & $<\log(R)>$ & $\sigma_{R}$\\ 
\hline
{\it Total subsamples} & & & & & & &  \\
AMIGA (all) & 207 & 9.80$\pm$0.05 & \_  & 28 &9.16$\pm$0.09 & \_ & $-$0.56$\pm$0.03 & \_ \\
AMIGA (det.) & 179 & 9.83$\pm$0.04 & 0.56 & 0 &9.38$\pm$0.05 & 0.71 & $-$0.45$\pm$0.02 & 0.31 \\
CfA (TS92) (all)  & 1544 & 9.89$\pm$0.01 & \_   & 210 & 9.42$\pm$0.02 & \_ & $-$0.44$\pm$0.02 & \_ \\
CfA (TS92) (det.) & 1334 & 9.89$\pm$0.01 & 0.54 & 0 & 9.59$\pm$0.02 & 0.73 & $-$0.31$\pm$0.01 & 0.41 \\
\hline
{\it Galaxies in common} & & & & & & &  \\
AMIGA(all) & 98 & 9.82$\pm$0.06 & \_   &  4 & 9.37$\pm$0.07 & \_   & $-$0.43$\pm$0.03 & \_   \\
AMIGA (det.)& 87 & 9.84$\pm$0.06 & 0.55 &  0 & 9.44$\pm$0.07 & 0.66 & $-$0.40$\pm$0.03 & 0.29 \\
CfA (TS92) (all)& 98   & 9.85$\pm$0.05 & \_   &  7 & 9.37$\pm$0.07 & \_   & $-$0.47$\pm$0.04 & \_   \\
CfA (TS92) (det.)& 87   & 9.85$\pm$0.05 & 0.50 &  0 & 9.43$\pm$0.07 & 0.62 & $-$0.42$\pm$0.04 & 0.33 \\
\hline
\end{tabular}

The entries are: 
{\it Column 1}: Sample considered. Both samples are selected with the same 
magnitude limit of (uncorrected) $m_{\rm zw} \le 14.5$. 
The distances of the galaxies of the CfA
sample have been derived in the same way as for the AMIGA sample 
(see Sect.~\ref{lfir_lb_comparison}). 
We give both the results obtained with ASURV (first row)
and the results with detections only (second row). 
{\it Column 2}: Total number of galaxies.
{\it Column 3}: Mean value of $\log(L_{B})$ and its error.
{\it Column 4}: Standard deviation of  $\log(L_{B})$.
{\it Column 5}: Number of galaxies with upper limits  in $L_{\rm FIR}$.
{\it Column 6}: Mean value of $\log(L_{\rm FIR})$ and its error.
{\it Column 7}: Standard deviation of  $\log(L_{\rm FIR})$.
{\it Column 8}: Mean value of $R= \log(L_{\rm FIR}/L_{B})$ and its error.
{\it Column 9}: Standard deviation of  $\log(L_{\rm FIR}/L_{B})$.
\label{comp_sauvage}
\end{table*}

\begin{figure}
\resizebox{1.0\hsize}{!}{\includegraphics{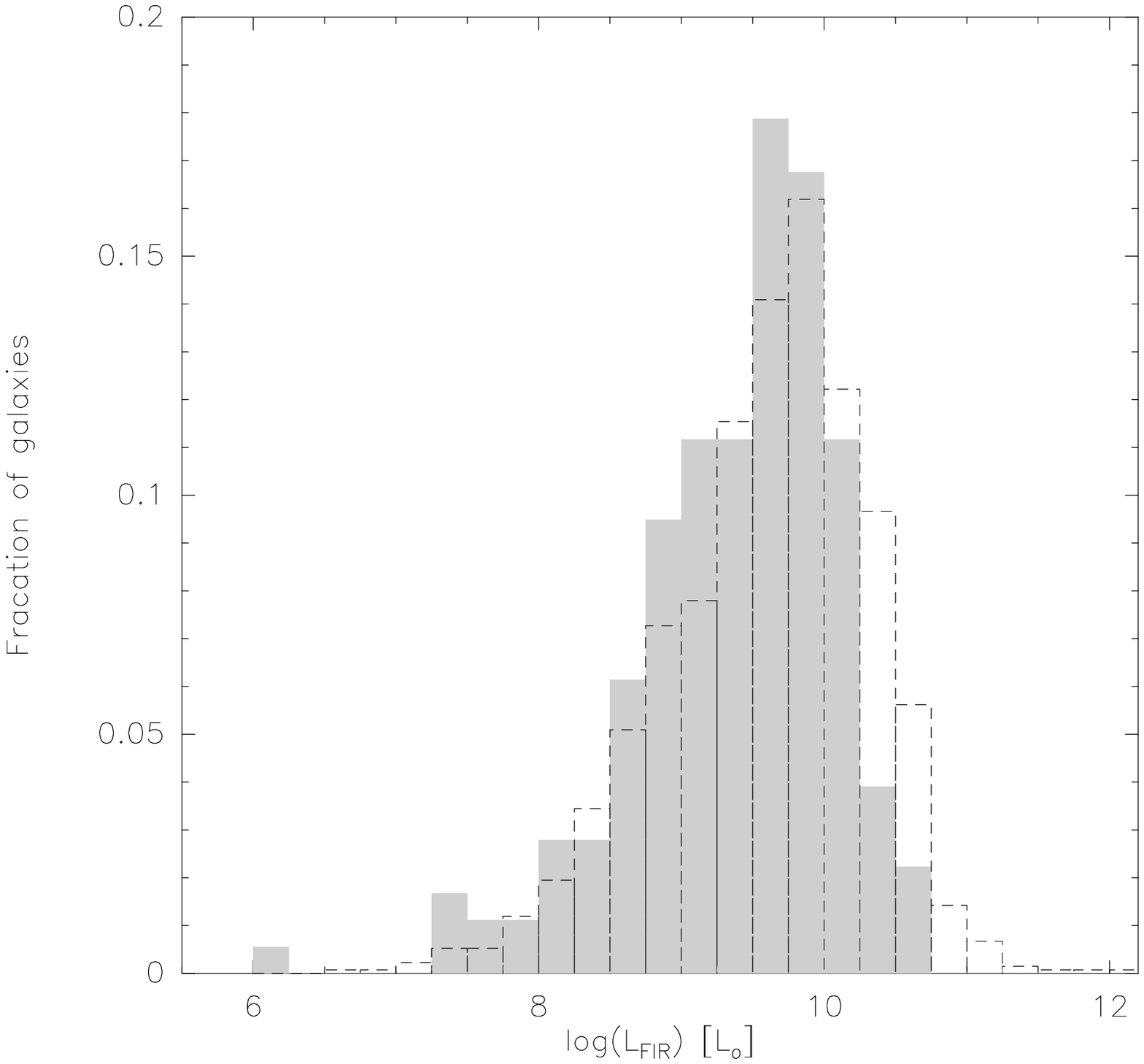}}
\caption{ The percentage FIR luminosity distribution for the FIR detections 
in the AMIGA sample restricted to (uncorrected) $m_{\rm zw} \le 14.5$  
(shaded area) and the corresponding distribution for the CfA sample 
(dotted line).}
\label{lfir_comp_sauvage_histo}
\end{figure}

In Fig.~\ref{lfir_comp_sauvage_histo} we show a comparison of our distribution
of $\log(L_{\rm FIR})$ to that of the CfA sample.  
Above $\log(L_{\rm FIR}/L_{\sun}) = 10.2$  a clear 
excess of CfA galaxies in comparison to our sample is visible.
In Table~\ref{comp_sauvage} we list the mean values. 
The difference between the mean value of  $\log(L_{\rm FIR})$ of the AMIGA and
the CfA sample is 0.21--0.26 (with and without taking into account upper limits)
which is a difference of 3--4$\sigma$.
We performed statistical two-sample tests  in the package ASURV and found
that the two distributions were different with a probability between
97.22\% (Logrank test) and 99.87\% (Gehan's Generalised Wilcoxon Test).
The maximum probability increases to $>$ 99.95 \% when only 
detections are taken into account.
We also performed  a Kolmogoroff-Smirnoff test on the detected data points
and derived a probability of more than 99.75\% that  the mean values
of $L_{\rm FIR}$ are different.
Therefore, there is strong statistical evidence that the AMIGA sample has
a lower $L_{\rm FIR}$ than the CfA sample, which is comparable but 
not selected with respect to the environment. 
This suggests that the FIR luminosity is a variable driven by interaction.

\begin{figure}
\resizebox{1.0\hsize}{!}{\includegraphics{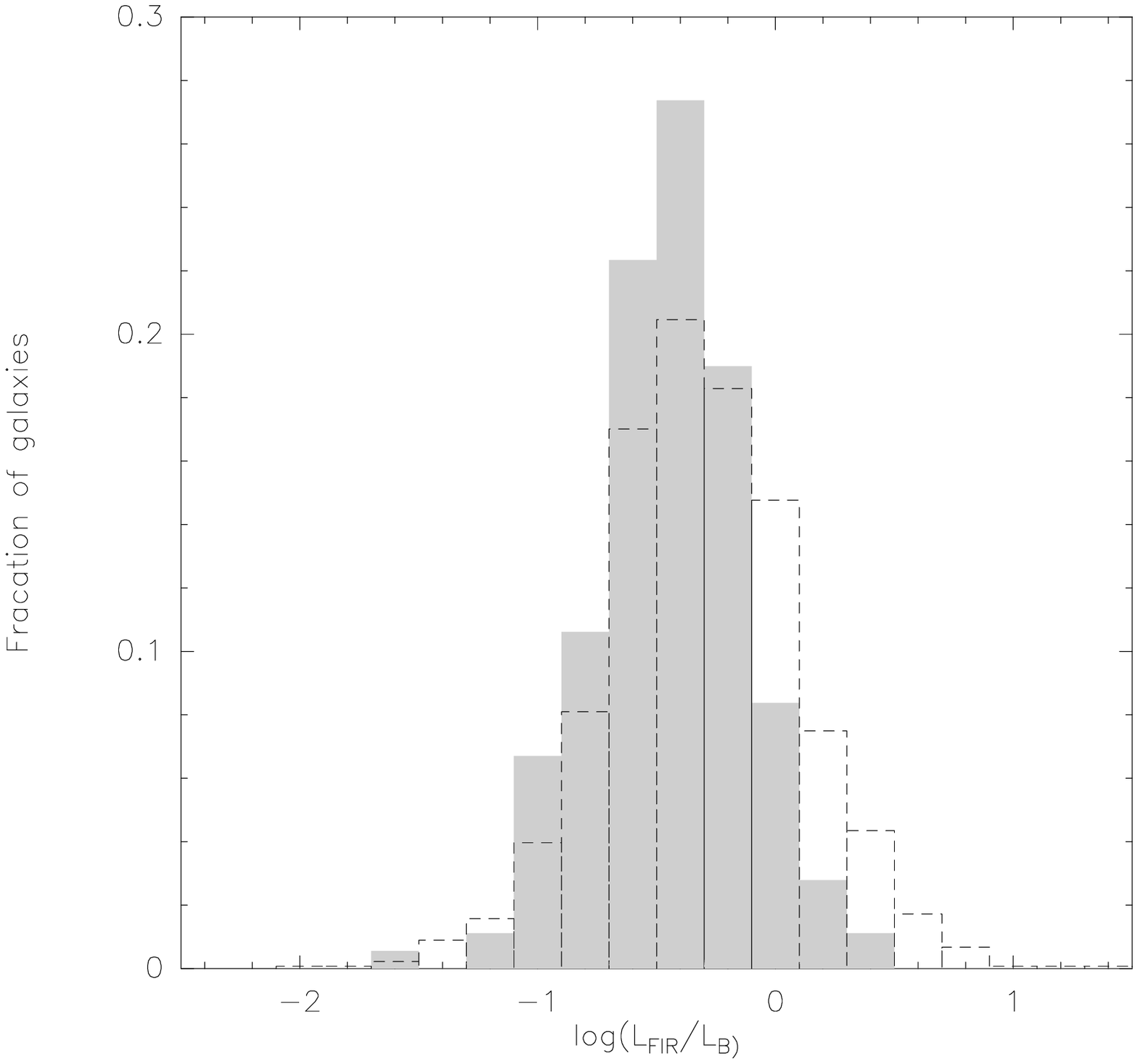}}
\caption{The percentage distribution of $R= \log(L_{\rm FIR}/L_{B})$ 
for the FIR detections in the 
AMIGA sample restricted to (uncorrected) $m_{\rm zw} \le 14.5$  
(shaded area) and the corresponding distribution for the CfA sample 
(dotted line).
}
\label{lfir_lb_comp_sauvage_histo}
\end{figure}

The comparison of the distribution of $R$ is shown in
Fig.~\ref{lfir_lb_comp_sauvage_histo}. 
We notice that the mean value of $R$  is higher for the CfA sample 
than for the AMIGA sample.  The difference is
0.12--0.14 (with and without upper limits) (see Table~\ref{comp_sauvage}) 
which corresponds to 4--7$\sigma$.
This difference has its origin in the higher value for $L_{\rm FIR}$ of the 
CfA sample, as the mean values for $L_{B}$\ are very similar
(see Table~\ref{comp_sauvage}), and the distribution of $L_{B}$ for both
samples (not shown here) practically agrees.
We performed the statistical two-sample tests  in the package ASURV and found
that the two distributions of $R$ were different with a probability between
97.3\% (Logrank test) and $>$ 99.95\% (Gehan's Generalised 
Wilcoxon Test).
On the other hand, the tests showed that the distributions of $L_{B}$\ were
identical with a non-negligible probability (28--68\%) confirming  that the 
difference in $R$ has its origin in $L_{\rm FIR}$.
Performing a Kolmogoroff-Smirnoff test on the detected data 
confirmed these results, yielding a probability of more than 99.999\%
that the mean values of $R$ are different.

\subsubsection{IRAS colours}
We compared the distribution of the IRAS colours to the results found for
the IRAS Bright Galaxy Sample (BGS, Sanders et al. \cite{sanders03}).
The value of $\log(F_{\rm 60}/F_{\rm 100})$  of the 
BGS is higher by about
0.2--0.3 with respect to our sample. This is not surprising, since the BGS
contains galaxies in a more active star forming phase than the CIGs.
The peaks of the distribution of the other colours in the BGS are very similar 
to ours, the only exception being the 
asymmetric tail towards high values of $F_{\rm 25}/F_{\rm 60}$  
which is absent in the BGS. 

\begin{table*}
      \caption{Comparison of IRAS colours to other studies.}
\begin{tabular}{lcccccc}
\hline
\hline
(1) & (2) & (3) & (4) & (5) & (6) & (7)  \\
Sample & $n/n_{\rm up}$ & $<\log(\frac{F_{\rm 60}}{F_{\rm 100}})>$ &  
$n/n_{\rm up}$& $<\log(\frac{F_{\rm 25}}{F_{\rm 60}})>$  & $n/n_{\rm up}$
 & $<\log(\frac{F_{\rm 12}}{F_{\rm 25}})>$ \\
\hline
{\it Isolated samples}  \\
AMIGA total                  & 468/76  &$-$0.42$\pm$0.01 & 468/343 &$-$0.87$\pm$0.02 & 126/67 &$-$0.33$\pm$0.03  \\
AMIGA, only det.             & 392/0   &$-$0.44$\pm$0.01 & 125/0 &$-$0.76$\pm$0.02 & 59/0 &$-$0.18$\pm$0.03  \\
XS91 CIG                     & 261/\_ &$-$0.42$\pm$0.01 & \_& $-$0.96$\pm$0.02 &\_ &$-$0.32$\pm$0.04 \\
Bushouse et al. (\cite{bushouse88}) isolated& 68/0& $-$0.39 & \_ &\_ & 34/0 & $-$0.21 \\
AMIGA m$_{\rm zw(uncorr)}<14.5$ & 183/4   &$-$0.43$\pm$0.01 & 183/90 &$-$0.88$\pm$0.02 & 93/35 &$-$0.26$\pm$0.03  \\
AMIGA m$_{\rm zw(uncorr)}<14.5$, (det.)& 179/0 &$-$0.43$\pm$0.01 & 93/0 &$-$0.82$\pm$0.02 & 58/0 &$-$0.18$\pm$0.02  \\
CfA (ST92)      & 1465/131 & $-$0.42$\pm$0.004 & 1465/771 & $-$0.94$\pm$0.01 & 706/154 & $-$0.26$\pm$0.01 \\
CfA (ST92), (det) & 1330/0   & $-$0.43$\pm$0.004 & 694/0& $-$0.87$\pm$0.01 & 552/0 & $-$0.21$\pm$0.01 \\
\hline
{\it Interacting samples} \\
AMIGA interacting            & 14/2 &$-$0.36$\pm$0.03 & 14/10 & $-$0.87$\pm$0.03 & 4/0 & $-$0.32$\pm$0.08 \\
XS91 wide pairs              & \_ &  $-$0.39$\pm$0.01&\_ & $-$0.93$\pm$0.02& \_& $-$0.52$\pm$0.07 \\
XS91 close pairs             & \_ & $-$0.31$\pm$0.01 &\_ &  $-$0.93$\pm$0.02&\_ & $-$0.65$\pm$0.08 \\
Bushouse et al. (\cite{bushouse88}) inter.& 98/0  & $-$0.33 &\_ &\_ & 48/0 & $-$0.42 \\
\hline
\end{tabular}

The entries are: 
{\it Column 1}: Sample considered.
{\it Columns 2, 4, 6}: Total number of galaxies and number of galaxies 
with upper limits. 
{\it Columns 3, 5, 7}: IRAS colour. For ratios involving $F_{\rm 60}$, 
only data points with detections at this wavelength were considered, whereas 
in  $<\log(F_{\rm 12}/F_{\rm 25})>$ only data points with detections at 
25~$\mu$m were taken into account. A ``\_'' means that the corresponding 
data point was not given in the reference. 
\label{colors_comparison}
\end{table*}

A comparison to the results of XS91 for a smaller subsample of 
CIG galaxies (see Table~\ref{colors_comparison}) shows an 
excellent agreement of the values for $\log(F_{\rm 60}/F_{\rm 100})$ and 
$\log(F_{\rm 12}/F_{\rm 25})$. Our value for 
$\log(F_{\rm 25}/F_{\rm 60})$ is however slightly  higher than that of XS91.
We also compared our results to the CfA sample studied by ST92. For this aim,
we produced again a different subsample, carrying out
the same magnitude cut ($<$ 14.5 in uncorrected Zwicky magnitude) as
in ST92. We found a very good agreement for $\log(F_{\rm 60}/F_{\rm 100})$ 
(see Table~\ref{colors_comparison}), and for $\log(F_{\rm 12}/F_{\rm 25})$.
With respect to $\log(F_{\rm 25}/F_{\rm 60})$, we derived a slightly higher 
value for our sample. However, we consider the significance of this difference  low
due to the large  number of upper limits.

ST92 found in their  analysis the same trend with Hubble type for  
$\log(F_{\rm 60}/F_{\rm 100})$ as we did. The value that they found for ellipticals,  
$\log(F_{\rm 60}/F_{\rm 100}) = -0.38$, 
is slightly higher than ours whereas their value for
irregulars, $\log(F_{\rm 60}/F_{\rm 100}) = -0.32$, agrees very well.
Also their values for spirals (between $-0.45$ and  $-0.47$ for $T = 2$--5)
are very close to ours. They explained the high
$F_{\rm 60}/F_{\rm 100}$ ratio in ellipticals by the 
concentration of the dust in the central regions where the radiation field 
is high, producing in this way a higher dust temperature.
A high $F_{\rm 60}/F_{\rm 100}$ ratio for irregulars has been
found in other studies as well (e.g. Melisse \& Israel \cite{melisse94}) and 
can be understood as a lack of ``cirrus'' emission with respect to 
FIR emission from H\,{\sc ii} regions.

\subsection{Comparison to interacting galaxies}

\subsubsection{$L_{\rm FIR}$\ and $L_{B}$}
One of the motivations for refining and studying a large 
sample of isolated galaxies is to better define a baseline against which effects of
environment could be quantified. Both mean IR diagnostic measures and
their dispersion are of interest in this context. AMIGA began with a CIG
sample selected to avoid as much as possible near neighbours.
Yet visual reevaluation of the optical morphologies for the sample
using POSS2/SDSS revealed 32 objects with clear signs of interaction
(Paper~II). These galaxies have been excluded from the present effort
to characterise the isolated sample but offer a useful internal
comparison sample to measure the sensitivity of the IR diagnostics to
environment. Fourteen of these 32 galaxies have IRAS  measures and apparent
magnitudes between 11 and 15. Table~\ref{tab_average} 
shows that the mean $\log(L_{B})$ is almost
identical to the isolated sample while $\log(L_{\rm FIR})$ is brighter by 
$\sim$0.7 (respectively 0.6 when comparing to the spiral/irregular subsample).
Similarly the mean FIR-to-optical flux ratio,
$<R>$, is 0.50 (respectively 0.43 for the spiral/irregular subsample)
higher. 
This  shows that there is  significant difference in $R$ between isolated 
and interacting galaxies due to an enhancement in $L_{\rm FIR}$\ of the latter.

\begin{figure}
\resizebox{1.0\hsize}{!}{\includegraphics{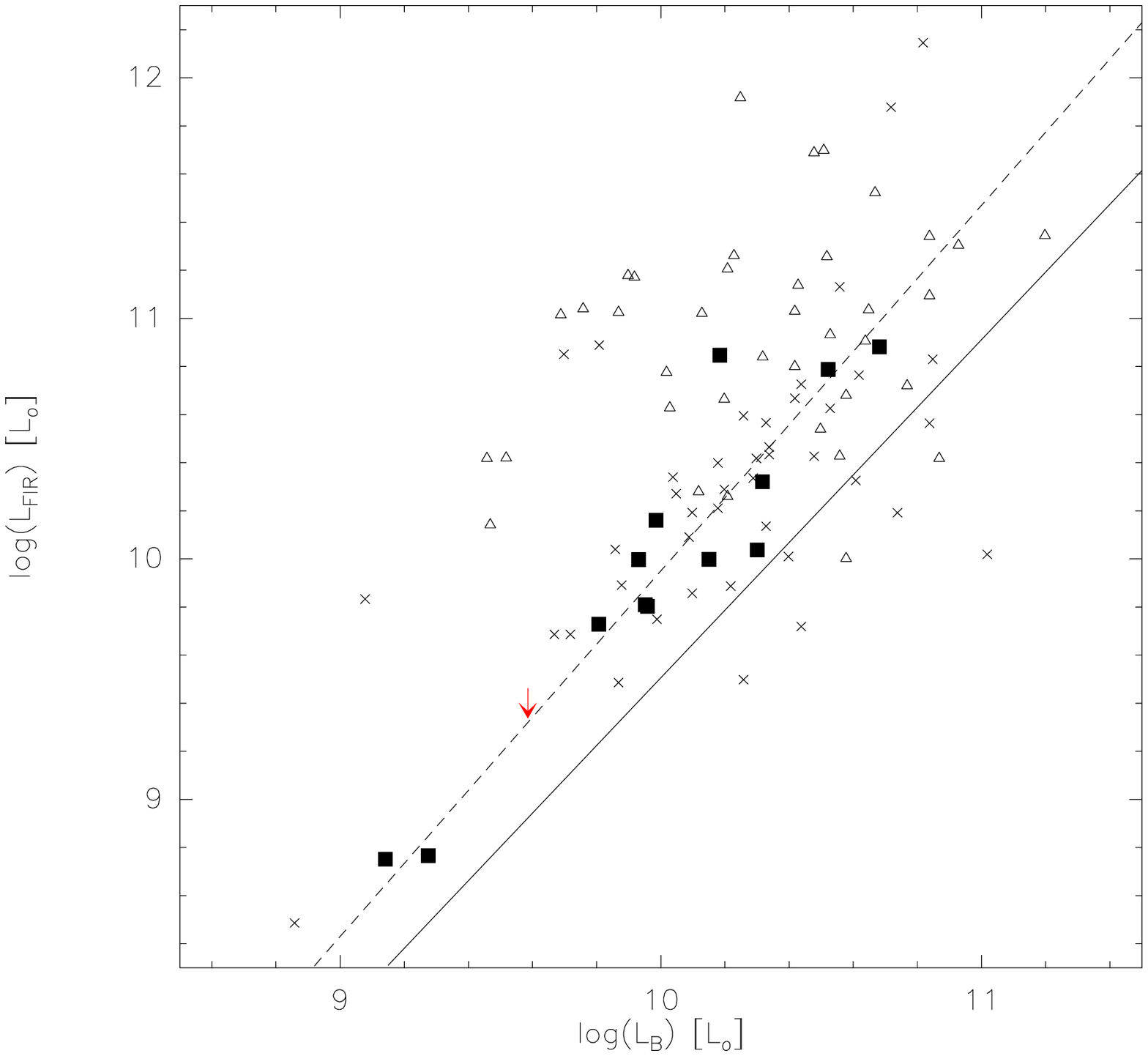}}
\caption{
$L_{\rm FIR}$ vs. $L_{B}$ for various interacting subsamples. 
The filled squares  and arrows
denote the galaxies from the CIG showing signs of interaction (see Paper~II).
The dashed line is the regression fit to this
subsample. 
The triangles indicate strongly interacting galaxies,
and the crosses weakly interacting galaxies from Perea et al. (\cite{perea97}).
The full line is the fit to the total AMIGA sample ($n=701$) of 
Fig.~\ref{lfir_lb_all}.
}
\label{lfir_lb_inter}
\end{figure}

Figure~\ref{lfir_lb_inter} shows the correlation between $L_{\rm FIR}$ and 
$L_{B}$ for this interacting subsample. Regression analysis yields a steeper
slope (see Table~\ref{regression}), as is usually found for interacting
samples, indicating that $L_{\rm FIR}$\ increases faster as a function of
$L_{B}$\ in comparison to the non-interacting sample.
The reason for this FIR excess is most likely an enhancement of
(dust-enshrouded) SF in interacting galaxies.
This is consistent with the results
in Perea et al. (\cite{perea97}) where a FIR enhancement was found for a sample
of perturbed galaxies.  We included their weakly and strongly
interacting samples in Fig.~\ref{lfir_lb_inter}. The effect is
strongest for their strongly perturbed sample.  The average FIR excess
(i.e. average deviation from the AMIGA regression line) for the strong
and weak interacting samples are 0.87 and 0.49 respectively.

XS91 compared mean FIR-to-optical flux ratios and found
a much smaller, but significant, 
difference between spiral-spiral pairs ($R = -0.17$) and a late-type
subsample from the CIG ($R = -0.30$).  The difference increased when
considering only close pairs ($R = -0.02$).
In order to compare their result to ours, we have to take into account that
they used {\it uncorrected} Zwicky magnitudes. The correction that we applied 
(see Paper~I) 
changed the Zwicky magnitudes by on average $-$0.67 magnitudes, corresponding
to a change in $R$ of $-$0.27 dex. Taking this into account, the value 
$R$ for the CIG subsample of XS91 is practically the same ours, whereas
$R$ for the pair sample in XS91 is  below ($\sim$0.2--0.4) the value of 
our interacting sample. Due to the small size of our 
interacting sample and the different selection
(the study of XS91 restricted the environmental signature to the effects of
one-on-one interactions) we do not want to draw any conclusions from
this difference.

\subsubsection{IRAS colours}

We  found a slightly higher value of $\log(F_{\rm 60}/F_{\rm 100})$
for the possibly interacting galaxies in the CIG   than for the 
total AMIGA sample
(see Table~\ref{tab_colors}). The difference is, however, only 2$\sigma$, 
and thus not statistically significant.
In the other colours ($\log(F_{\rm 25}/F_{\rm 60})$, $\log(F_{\rm 12}/F_{\rm 25})$ or
$\log(F_{\rm 12}/F_{\rm 60})$) we found, within the errors, no difference between both samples. 
%

A trend of higher  $\log(F_{\rm 60}/F_{\rm 100})$  values in interacting 
galaxies has been found in other studies (see Table~\ref{colors_comparison}). 
XS91 compared their subsample of  CIGs to  paired galaxies. 
They found a value very close to our interacting sample for wide  late-type pairs
(i.e. distance between partners larger than
the diameter of the primary) and an even higher value, significantly higher
than for the value  for the CIGs, for  
close late-type pairs (i.e. of distance between partners less than
the diameter of the primary). Bushouse et al. (\cite{bushouse88}) studied 
the MIR/FIR properties of a sample of 109 colliding galaxies and compared
it to more isolated galaxies from the sample of 
Kennicutt \& Kent (\cite{kennicutt83}). In their study, they only took 
into account IRAS detections. 
Their values for  $\log(F_{\rm 60}/F_{\rm 100})$, both for the interacting
as well as for the comparison sample, is in reasonable agreement
with our study and that of XS91.

We did not find a significant difference in neither $F_{\rm 25}/F_{\rm 60}$ 
nor $F_{\rm 12}/F_{\rm 25}$ between isolated and interacting galaxies.
In contrast to this, XS91 obtained a lower value for $F_{\rm 12}/F_{\rm 25}$
both for the close  and the wide pair samples. Also Bushouse 
et al. (\cite{bushouse88}) found a higher value for $F_{\rm 12}/F_{\rm 25}$
in the interacting sample, although their result has to be taken with
caution because only detections were included (thereby skewing the result
to higher values).
Given the very low number of galaxies  with detections in our 
interacting subsample, the significance of these differences is, however,
not statistically meaningful.

\subsection{Nature of the FIR brightest galaxies}

MIR/FIR measures have been found to be sensitive diagnostics of enhanced SF.
Since SF can be greatly enhanced by the presence of companions we 
can ask if any of the most FIR luminous galaxies in our sample are really 
isolated. Naturally, as discussed in Paper~II, we are limited in our 
ability to describe galaxy form and to
detect close companions by the quality of the available images. 
\begin{table}
      \caption{Galaxies with $\log(L_{\rm FIR}/L_{\sun})>10.5$}
\begin{tabular}{lccccccccccc}
\hline
\hline
(1) & (2) & (3) & (4)  \\
CIG  & $\log(L_{\rm FIR}/L_{\sun})$ & Hubble type & Comment \\
\hline  
55  & 11.12  &     Sc ($T=5$)  &  I/A + Sy/LINER\\
143 & 10.86  &      Sbc   ($T=4$)  & I/A? lopsided spiral\\
148 &  10.56  &      Sbc    ($T=4$)  & I/A\\
232  &  10.69 &       Sc    ($T=5$) & I/A? \\ 
302  &  11.02  &       Sc   ($T=5$) & I/A? peculiar\\
361  &  10.51  &       Sab   ($T=2$) & Isolated spiral?\\
444  &  10.54  &       Sb  ($T=3$) & Possibly Sm companion\\
709  &  10.53  &       Sc   ($T=5$) &  Sm companion nearby  \\
715  &  10.76  &       Sc   ($T=5$)&  I/A\\
829  &  10.51  &       Sb   ($T=3$)&   Blue Compact\\
841  &  10.58  &       S0   ($T=-2$) & large inclined S0 \\
866  &  10.95  &       Sb  ($T=3$)&  Isolated, LINER\\
\hline
\end{tabular}

The entries are: 
{\it Column 1}: CIG number. 
{\it Column 2}: FIR luminosity. 
{\it Column 3}: Hubble type. 
{\it Column 4}: Comment after visual inspection of optical images, and
consultation of NED.
I/A means interacting, I/A?: possibly interacting, Sy: Seyfert galaxy 
\label{lfir_high}
\end{table}
Table~\ref{lfir_high} lists the FIR brightest galaxies (with 
$\log(L_{\rm FIR}/L_{\sun})>10.5$) in our sample and the results of 
reexamination of POSS2 and other data sources for them. More than half of 
this sample are likely to involve interacting systems. A very few may 
represent the most luminous examples of self-induced star formation or 
the IR measures are dominated by a yet undetected active nucleus (many 
do not have published spectra). CIG~709 epitomises another issue raised in 
Paper~II. While it appears to be isolated from similarly sized objects, 
it shows striking structural asymmetries and a small companion about one 
diameter distant. It is yet unclear whether such small companions are 
capable of significantly enhancing star formation and producing structural 
deformations in massive spirals (see also Espada et al. \cite{espada05}). 

We furthermore inspected the most extreme outliers from the regression fit
between $L_{\rm FIR}$\ and $L_{B}$. 
The outliers on the high side come under immediate suspicion as interactors
that were missed in the morphology survey. Only two show 
$\log(L_{\rm FIR})>11.0$ and both are possible interacting systems -- in 
NED they are described as Seyfert/LINER (CIG~55) and H\,{\sc ii} galaxy (CIG~302). 
Several others are  candidate interactors or have an
active nucleus. Galaxies falling below the regression are either early-types
(E and S0), known to be deficient in FIR emission or highly
inclined galaxies. The internal extinction correction for such objects
is large and uncertain suggesting that the low $L_{\rm FIR}$/$L_{B}$ 
measures may be due to overcorrection of the blue magnitude.


\section{Conclusions\label{section6}}

We present a MIR/FIR analysis of a sample of the most isolated
galaxies in the local universe, obtained from the Catalogue of Isolated
Galaxies (CIG). This analysis is  part of our 
multiwavelength study of the properties of the interstellar medium
of this sample and involves
ADDSCAN/SCANPI reprocessing of IRAS data for all 1030 galaxies (out of 1050 in 
the CIG) covered by IRAS. We increased the detection rate with respect to the 
PSC and FSC in all IRAS bands and present our AMIGA sample of 701 CIG 
galaxies as the best available control sample against which to evaluate the 
IR signatures of environment in local galaxy samples.
Our sample is large enough to permit comparison of IR properties 
for galaxy morphology subclasses. Our main results are the following:
\begin{enumerate}
\item  The galaxies in our sample have  modest FIR luminosities,
with only 14 objects (corresponding to $<$2\% of the
sample) above $\log(L_{\rm FIR}/L_{\sun}) = 10.5$. 
The mean $\log(L_{\rm FIR})$ of our sample is 3--4$\sigma$ (0.21--0.26 dex)
below the corresponding value for the CfA sample
studied by TS92 and ST92, which is
a sample of nearby galaxies similar to ours, but selected without
considering the environment. In addition, a lower value
(by 0.12--0.14 dex, corresponding to 4--7$\sigma$), compared to the CfA 
sample, was found for the mean $R=\log(L_{\rm FIR}/L_{B})$ of our sample. 
This suggests that the FIR emission is
a parameter driven by interaction and that our sample of 
isolated galaxies shows a value close to the lowest possible.

\item We find evidence for a systematic increase 
in FIR luminosity from type S0/a to Sc followed by a decline for later types
(dominated by dwarf galaxies), 
possibly reflecting lower dust masses in those galaxies or less efficient 
star formation. 
At the same time, $R$ is essentially constant for all Hubble types later
than S0/a 
suggesting that the SF efficiency in isolated 
spirals and irregulars is roughly constant.

\item Early-type galaxies (E and S0) show a lower average
$R$ than the spirals. We can divide them into two
populations: 1) undetected galaxies (the majority) with low upper
limits in $R$ and 2) a population of early type with $L_{\rm FIR}$ and $R$ 
that are similar to spiral galaxies. The latter galaxies require confirmation 
of the assigned early-types. If real, they represent  an interesting class
of isolated galaxies.

\item We calculated the bivariate FIR luminosity function which was
found to be in good agreement
with previous studies  (XS91) based on a smaller subsample of the CIG.
The FIRLF is dominated by  moderately FIR luminous galaxies (only 3 objects
have $\log(L_{\rm FIR}/L_{\sun}) > 11.0$) and is well described by a Schechter 
function. This contrasts to results obtained for FIR selected samples 
(e.g. Soifer et al. \cite{soifer86}; Sanders et al. \cite{sanders03}) where 
a double power law is needed to achieve a fit to the high-luminosity end
of the FIRLF. 

\item We found a correlation between $L_{\rm FIR}$ and $L_{B}$ with a
slope  above 1 (\nobreak{$L_{\rm FIR}$ $\propto$ $L_{B}$$^{1.41}$}) with
only modest variations as a function of Hubble type. 
Possible reasons for the slope being $>$1 are an increase in extinction or
an enhancement of the recent  SF activity  with galaxy luminosity.
%

\item We identified a small population of possibly interacting galaxies in 
the CIG (Paper~II) and these show a significantly higher mean FIR luminosity
than the rest of the sample.  
They  lie above the regression line
derived for our optically selected CIG sample.  The same was found for
samples of interacting galaxies from Perea et al. (\cite{perea97}).
This suggests that the FIR emission is enhanced due to the interacting.

\item We found
a trend of $F_{\rm 60}/F_{\rm 100}$
with Hubble type: elliptical galaxies and irregular galaxies have a
higher $F_{\rm 60}/F_{\rm 100}$ than spirals, indicating a hotter dust 
temperature. For the ellipticals the most likely reason is
the higher concentration of dust towards the inner regions of the
galaxies (ST92) whereas in irregulars, a lack of
cirrus emission is the most probable cause.

\item The value of  $F_{\rm 60}/F_{\rm 100}$ of the AMIGA sample was found to
be lower than that of interacting samples from the literature
(XS91 and Bushouse  \cite{bushouse88}) indicating that interaction
can increase the dust temperature.

\end{enumerate}

As the largest and most isolated sample in the local Universe, AMIGA
can serve as a valuable control when assessing the effects of
environment on other local samples of galaxies. This can in turn 
clarify our interpretation of the FIR signature in samples at higher redshift.

\begin{acknowledgements}
{We would like to thank M. Sauvage for making the data of his sample
available to us.
This research has made use of the 
NASA/IPAC 
Extragalactic Database (NED) which is 
operated by the Jet Propulsion Laboratory, California Institute of Technology, 
under contract with the National Aeronautics and Space Administration,
and of the Lyon Extragalactic Database (LEDA).
This work has been partially supported by DGI Grant
AYA 2005-07516-C02-01 and the Junta de Andaluc\'{\i}a (Spain).  
UL acknowledges support by the research project ESP\,2004-06870-C02-02.
JS is partially supported by a sabatical grant SAB2004-01-04 of 
the Spanish Ministerio de Educaci\'on y Ciencias.
GB acknowledges support at the IAA/CSIC by an I3P contract (I3P-PC2005F) 
funded by the European Social Fund.}
\end{acknowledgements}














\end{document}